\documentclass[titlepage,11pt,twoside]{article}

\usepackage[dvips]{graphicx}

\usepackage[myheadings]{fullpage}
\usepackage{pmetrika}
\usepackage{pmbib}

\usepackage{tikz}
\usepackage{amssymb,amsmath,amsfonts,amsbsy}
\usepackage{graphicx}
\usepackage{numprint}

\tikzstyle{ov}=[shape=rectangle,draw=black!80,minimum height=0.6cm,
                minimum width=0.6cm,thick,fill=white]
\tikzstyle{lv}=[shape=circle,draw=black!80,thick,minimum width=0.7cm,fill=white]
\tikzstyle{lr}=[shape=circle,draw=black!00,thick,minimum width=0.7cm]
\tikzstyle{pt}=[shape=circle,draw=black!100,thick]

\usepackage{submit}

\begin{document}
\begin{titlepage}

\linespacing{1}

\title{Forecasting intra-individual changes of affective states taking into account inter-individual differences using intensive longitudinal data from a university student drop out study in math}

\author{Augustin Kelava}
\affil{\small{Methods Center, University of Tuebingen}}

\vspace{3ex}

\author{Pascal Kilian}
\affil{\small{Methods Center, University of Tuebingen}}

\vspace{3ex}

\author{Judith Glaesser}
\affil{\small{Methods Center and Tuebingen School of Education, University of Tuebingen}}

\vspace{3ex}

\author{Samuel Merk}
\affil{\small{Methods Center and Tuebingen School of Education, University of Tuebingen}}

\vspace{3ex}

\author{Holger Brandt}
\affil{\small{Department of Psychology, University of Zurich}}

\markboth{Psychometrika}{ }

\vspace{\fill}

\vspace{\fill}

\linespacing{1}\fontsize{8}{10}\selectfont




\end{titlepage}\vspace*{24pt}

\linespacing{1}

\RepeatTitle{Forecasting intra-individual changes of affective states taking into account inter-individual differences using intensive longitudinal data from a university student drop out study in math}


\begin{center}\vskip3pt

\vspace{32pt}

Abstract\vskip3pt

\end{center}

\begin{abstract}
The longitudinal process that leads to university student drop out in STEM subjects can be described by referring to a) inter-individual differences (e.g., cognitive abilities) as well as b) intra-individual changes (e.g., affective states), c) (unobserved) heterogeneity of trajectories, and d) time-dependent variables. Large dynamic latent variable model frameworks for intensive longitudinal data (ILD) have been proposed which are (partially) capable of simultaneously separating the complex data structures (e.g., DLCA; Asparouhov, Hamaker, \& Muthén, 2017; DSEM; Asparouhov, Hamaker, \& Muthén, 2018; NDLC-SEM, Kelava \& Brandt, 2019). From a methodological perspective, forecasting in dynamic frameworks allowing for real-time inferences on latent or observed variables based on ongoing data collection has not been an extensive research topic. From a practical perspective, there has been no empirical study on student drop out in math that integrates ILD, dynamic frameworks, and forecasting of critical states of the individuals allowing for real-time interventions. In this paper, we show how Bayesian forecasting of multivariate intra-individual variables and time-dependent class membership of individuals (affective states) can be performed in these dynamic frameworks. To illustrate our approach, we use an empirical example where we apply forecasting methodology to ILD from a large university student drop out study in math with multivariate observations collected over 50 measurement occasions from multiple students (N = 122). More specifically, we forecast emotions and behavior related to drop out. This allows us to model (i) just-in-time interventions, (ii) detection of heterogeneity in trajectories, and (iii) prediction of emerging dynamic states (e.g. critical stress levels or pre-decisional states).\vskip3pt

\end{abstract}

\vspace{\fill}\newpage

\section{Introduction}
In this paper, we introduce an innovative methodological approach set in a dynamic framework which can be used to forecast critical states, allowing for real-time inferences on latent or observed variables based on ongoing data collection. It is particularly well suited to analysing intensive longitudinal data (ILD). After introducing the model and the Forward Filtering Backward Sampling method (FFBS) the forecasting relies on, we apply it to the substantive area of university student drop out in mathematics, illustrating the approach using an empirical example. In a brief simulation study, we examine the forecasting performance of the model.

Forecasting, explaining, and preventing university student drop out from science, technology, engineering, and mathematics (STEM) subjects is an important issue both for economies and individuals. Drawing on the literature in the field, we identify gaps in existing knowledge and suggest that our methodologically innovative approach based on intensive longitudinal data (ILD) is particularly well suited to studying potential contributing factors to university student drop out. In the brief literature review below, we first describe some classic explanatory models of student dropout, followed by studies focusing on possible interactions between contributing factors. All these suggest the need for a process perspective, given the fact that the intention to drop out evolves over a certain period of time; in other words, it is a process rather than an event at a single point in time. 

\subsection{University student drop out from math}
STEM subjects have the highest rates of student drop out and the first semester is the critical phase in STEM studies (e.g., Heublein, 2014; Heublein et al., 2017, for the German context; see also Burrus et al., 2013; Robbins et al., 2004; Witteveen \& Attevell, 2017). This finding together with the importance of STEM subjects in modern economies makes the study of drop out in math particularly urgent. There is a large body of research on the topic, though various gaps remain which the current study aims to address.

Two classic models of university drop out are Tinto’s model of student departure (Tinto, 1993) and, complementing it, Bean’s model of student attrition (Bean, 1980, 1983). In their review of the literature, Burrus et al. (2013) note that many studies are based on these two models. Factors commonly studied can be grouped into eight areas: (a) institutional environment factors, (b) student demographic characteristics, (c) commitment, (d) academic preparation and success factors, (e) psychosocial and study skills factors, (f) integration and fit, (g) student finances, and (h) environmental pull factors. Given the large number of factors which might be associated with drop out and drop out intentions, a choice of focus on either institutional/contextual or individual factors or the interaction of these has to be made by researchers. The present study focusses on individual factors which will be discussed in the next section.

\subsection{Attainment, motivation, affect: Main effects and interaction effects}
One of these individual factors is motivation. Not surprisingly, it has been shown to be involved in student drop out. Dresel and Grassinger (2013) show that both motivational state measured at the beginning of a course of study and changes in motivation over the course of the first semester have an effect on the intention to drop out or to change course.

Ghassemi et al. (2017) frame university drop out as a goal attainment issue. They use longitudinal data and multilevel modelling techniques to investigate the role of “action crisis” which is “operationalized by individuals’ reports of several facets constitutive of conflict” (p. 525), i.e. motivational measures such as value expectancy and (psychological) cost measures. The authors did indeed find a link between such measures and psychological processes leading to university drop out. Several meta-analyses also showed that non-cognitive factors such as motivational and emotional measures are related to the outcomes university success and retention (e.g., Richardson et al., 2012; Robbins et al., 2004). Richardson et al. (2012) note in their theoretical introduction that self-regulatory learning strategies may moderate the effects of dispositional characteristics such as intellectual capacity and personality (p. 360), but in their discussion they point out that there is a need for more research into possible moderators.

Other studies also suggest that the investigation of interaction effects ought to be considered. Dresel and Grassinger (2013) investigate possible interaction effects between pre-university achievement and motivation on the outcome of meta-cognitive strategies. However, empirically they found that while the coefficient in the regression analysis was statistically significant, variance explained did not increase substantially. Similarly, Schnettler, Bobe, Scheunemann, Fries, and Grunschel (2020) note in their introduction that there might be “cross-level interactions (e.g., does the relative strength of the association between success expectation and drop out intention depend on person characteristics such as gender?)” (p.497). Empirically, they only found an interaction for age, but not for gender, GPA, major or semesters studied, but they note in their discussion that further research on such interactions would be welcome.

\subsection{A process perspective}
Several authors (including some of those referred to in the previous section) discuss the processual nature of the decision to drop out of university. It seems that university drop out is best conceived of as the result of a process rather than a single sudden event which suggests that a longitudinal study design is appropriate for studying it (e.g., Dresel \& Grassinger, 2013; Ghassemi et al., 2017; Isphording \& Wozny 2018; Schaeper, 2019; Schnettler et al., 2020; Witteveen \& Attewell, 2017). The duration and frequency of data collection varies between studies, with data collection taking place no more than two or three times per semester. By contrast, Dietrich et al. (2017) undertake a study based on intensive longitudinal data (ILD), with data collection having taken place at a much higher frequency given the possibility that students react to events with changes in their motivational and emotional states. However, the authors' focus is not on drop out (to our knowledge, no such study exist on drop out specifically), though the outcome they study, effort, has some relevance to university drop out. 
Similarly, while not conducting an ILD, Witteveen and Attewell (2017) analyze longitudinal transcript data to examine processes leading to graduation, using Hidden Markov modeling (HMM). They identify several latent states that are associated with patterns of course taking, and show that HMM can predict graduation or nongraduation based on a few semesters of transcript data. 

The present study builds on this body of research, using an ILD approach in order to study university drop out and its precursor, the intention to drop out.
Individual factors potentially linked to university drop out, the focus of the present study, include a wide variety of characteristics which can be either fairly stable over time and across situations (e.g., Kilian, Loose, \& Kelava, 2020) or \textit{changeable within an individual over time}. The latter are the reason for taking a longitudinal perspective, as outlined in this section. Both Dietrich et al. (2017) and Schnettler et al. (2020) stress the importance of investigating such \textit{intra-individual changes} in addition to inter-individual differences since the latter may not capture individual psychological processes which have been shown to be linked to outcomes such as university drop out. Gender, cognitive ability and pre-university academic performance are examples of stable (inter-individual) characteristics, motivational and affective states as well as goal orientation are examples of states which can vary over time in response to external experiences and stimuli. The relevance of both types of factors for university drop out has been demonstrated in numerous studies, as we have seen.

\subsection{Implications}
Taken together, these findings point to the need for more research into possible interaction effects between factors contributing to university drop out. They also suggest that these factors may be located on two levels: one level comprises relatively stable personal characteristics or traits, the other consists of changeable psychological states. The investigation of a possible interaction between factors located on different levels appears particularly promising.

Given that factors on the state level can be volatile, they are best studied using a longitudinal design with high-frequency points of measurement given their changeable nature. Such a design would also make it possible to capture a third type of factor, time-dependent variables outside an individual’s control, i.e. external events which may influence (unobserved) heterogeneity of trajectories.
Accordingly, any study intending to capture this complex interplay of variables has to implement a design that is capable of mirroring the two levels on which factors related to student drop out are located and of studying possible interaction effects across levels. This methodological challenge requires innovative techniques so that the theoretical conceptualization can be captured by appropriate statistical models. To provide suitable data for the models, intensive repeated data collection is required. The next section outlines this paper’s aims arising from these issues.

\subsection{Aims of this article}
The aims of this article are as follows:
First, we give an overview on current large dynamic latent variable frameworks which address multivariate observations, unreliable measures (with unequal time intervals), and separate influences of the different types of data levels (within-person and across-person variability as well as time-dependent information and heterogeneity).

Second, we describe the data used in our empirical example. They were obtained as part of a large university math student drop out study with multivariate observations collected over time from multiple students. We describe the properties of the data set and study. Furthermore, we explain the dynamic model that we analyze and how it corresponds to substantive research questions as outlined in the introductory section.

Third, we describe the implementation of the Forecasting Filtering Backward Sampling (FFBS) procedure. We explain how the dynamic model can be used for the forecasting (e.g., of critical states). We give information on so-called observation and evolution equations, priors, state, forecast distributions, updating, and posterior probabilities which allow the prediction of states as well as the quantification of uncertainty.

The application of our procedure has two parts: (i) its use with a real data example and (ii) an examination of the robustness or our results via a small simulation study.
Therefore, fourth, we will apply forecasting methodology to the real data example. More specifically, we will forecast emotions and behavior (e.g., subjective experiences of overload, stress, positive and negative affective states, dysfunctional cognitions) related to drop out. Furthermore, we will analyze different influences that were proposed by substantive research which imply several data levels and complexity of models (including interactions). We will show the correspondence between theoretical expectations (see above) and empirical results.

Finally, in a small simulation study, we will examine the robustness of our results. We will address conditions that influence the stability of the empirical results produced by our forecasts.


\section{Dynamic factor and structural equation models}
In the past, dynamic factor (analysis) models were developed as techniques to describe intra-individual changes on latent variables (e.g., Molenaar 1985). Such models are (stationary) time series models with a factor-analytic structure. Over the past years, several extensions have been developed, for example, covering categorical variables (e.g., Zhang \& Nesselroade, 2007), Hidden Markov Models (e.g., Asparouhov \& Muthén, 2010; Hamaker \& Grasman, 2012; Hamaker, Grasman, \& Kamphuis, 2016), and continuous time perspectives, when data are sampled at unequal time intervals (e.g., Voelkle \& Oud, 2013; Oud, Voelkle, \& Driver, 2018).

In recent years, two-level extensions of dynamic factor models as described in Molenaar (1985), Zhang and Nesselroade (2007), or Zhang, Hamaker, and Nesselroade (2008) have been published. Large dynamic latent variable model frameworks have been proposed which are (partially) capable of simultaneously separating complex data levels (such as inter-individual differences, intra-individual changes, unobserved heterogeneity). For example, in the Dynamic Latent Class Analysis framework (DLCA; Asparouhov, Hamaker, \& Muthén, 2017), it is possible to specify heterogeneous autoregressive structural equation models. Intra-individual changes have class-specific patterns. Latent class membership follows a Hidden Markov process which is described by stable a priori inter-individual differences. Time-dependent information cannot be used to predict unobserved class membership (e.g., the status of intention to quit studies). 

Traditionally, in so called regime switching models, Hidden Markov processes (with latent classes) are used to explain heterogeneous autoregressive relationships (e.g., Chow \& Zhang, 2013; Dacco \& Satchell, 1999; Dolan, Schmittmann, Lubke, \& Neale, 2005; Kim \& Nelson, 1999; Hamaker \& Grasman, 2012; Tadjuidje, Ombao, \& Davis, 2009). The parameters on the intra-individual level vary depending on class membership. But class membership does not depend on the variables of the intra-individual level. For example, Chow and Zhang (2013) propose nonlinear regime-switching state-space models, which subsume regime switching nonlinear dynamic factor analysis models as a special case. The state-space processes are allowed to be nonlinear (within a so called regime). They model experience sampling data of positive and and negative affective states. In their approach, the transition probabilities are not explained by intra-individual changes. Dolan, Schmittmann, Lubke, and Neale (2005) apply regime switching to growth curve (mixture) modeling analyzing alcohol use data from the National Longitudinal Survey of Youth (NLSY97). The regimes are not influenced by intra-individual changes. As before, Hamaker and Grasman (2012) apply regime switching state-space models to psychological states associated with premenstrual syndrome, abrupt changes in affective states during major depression, and with the so called 'hot hand' phenomenon in sports. On the contrary, Tadjuidje, Ombao, and Davis (2009) propose a manifest vector autoregressive model (VARX) in which the regime switches depend on past values of the time series and exogenous variables. However, it is not a full latent variable approach.

In the Dynamic Structural Equation Model framework (DSEM; Asparouhov, Hamaker, \& Muthén, 2018), both time-dependent variables and inter-individual differences are used to explain random effects on the within level. However, unobserved heterogeneity (i.e., time-varying latent classes) is not part of the model. Multilevel dynamic models (e.g., Chow, Zu, Shifren, \& Zhang, 2011; Song \& Zhang, 2014) and their dynamic counterparts in continuous time (Oravecz, Tuerlinckx, \& Vanderkerckhove, 2009, 2016; Lu, Chow, Ram, \& Cole, 2019) have also been proposed in the literature. Chow, Zu, Shifren, and Zhang (2011) describe dynamic factor analysis models with time-varying parameters and fit daily affect data from participants with Parkinson’s disease. Their approach is classified as a multilevel dynamic model, but without latent classes/regimes. The same holds for Song and Zhang (2014), who describe the analysis of multivariate time series data of affect regulation and coregulation within couples using multilevel dynamic factor models. Oravecz, Tuerlinckx, and Vanderkerckhove (2009, 2016) present dynamic continuous time models as diffusion models for the analysis of affective states disentangling within- and between-person differences. Lu, Chow, Ram, and Cole (2019) propose a regime-switching dynamic continuous time model in which the regimes depend on inter-individual differences (latent variables and covariates), but not on intra-individual changes.

Although nonlinear latent effects were already considered in previous work (e.g., Chow, Tang, Yuan, Song, \& Zhu, 2011; Guo, Zhu, Chow, \& Ibrahim, 2012), a comprehensive approach capable of taking account of all properties had been missing. The Nonlinear Dynamic Latent Class Structural Equation Model framework (NDLC-SEM, Kelava \& Brandt, 2019) combines the capabilities of the frameworks and models mentioned above. It allows for the inclusion of time-dependent information (including intra-individual changes) for modeling unobserved time-dependent class membership and heterogeneity (e.g., unexpressed intentions to quit or affective states). Furthermore, stable inter-individual differences can also be used to explain changes between the unobserved time-dependent class memberships of individuals. On both the within and between level, nonlinear (semiparametric) effects can be specified and random effects are essential elements of the framework. In our analysis, we will use the key parts of the comprehensive NDLC-SEM framework, for example the capabilities to include nonlinear interaction effects on all levels as well as intra-individual changes (and inter-individual differences) that drive the latent class membership. We refer the reader to the original literature for an overview of all properties of the NDLC-SEM framework (Kelava \& Brandt, 2019).

\section{Intensive longitudinal data from the SAM study}
The empirical ILD used in this paper are taken from a large study on university student drop out in math. In this section, we describe the design of the study and the sample. Then we describe the model that we used for the forecasting of critical states related to drop out, and present the forecasting methodology.

\subsection{General setting of the study}
The SAM (German acronym for university drop out in mathematics) study is a longitudinal study focusing on university drop out in mathematics students at a German university. It is well documented that, in Germany, approximately 40 percent of students drop out in the early phase of math studies (Heublein, 2014). This is considerably higher than the 33 percent drop out rate found at German universities on average across all subjects in Heublein’s study. Internationally, rates vary considerably by country, with the OECD average being around 31 percent (OECD, 2010). The majority of students drop out during their first semester. Therefore, large introductory courses in math, such as calculus and algebra, are suitable settings to examine drop out in this important STEM subject. In the SAM study, data were collected from a first semester cohort starting in the winter semester of 2017/2018 and attending a calculus lecture and accompanying tutorial sessions. Since the calculus syllabi are very similar across German universities, the cohort can be considered to be prototypical for the general phenomenon of student drop out from math in Germany.

\subsection{Data collection}
The data is derived from three sources. The first source is a questionnaire used to obtain information on the level of potential inter-individual and contextual differences. Data was collected in particular on individual stable characteristics, such as scholastic performance, gender etc. during the second week of the first semester. Lecturers agreed to use a part of their lectures for students to complete the questionnaire. This contributed to achieving high participation rates ($N = 122$). Second, starting one week after the initial assessment, data on changeable individual affective and motivational states were collected to cover the level of potential intra-individual differences.
This was carried out via a short (5 minutes) online survey three times per week. The average participation rate per assessment day was $49.34$ ($SD = 17.72$) students, and on average the students participated in $20.22$ ($SD = 15.47$) out of 50 online surveys. 
Email invitations to the survey were sent out on Mondays, Wednesdays, and Fridays on a total of 50 measurement occasions (over a period of 131 days). Thus, it was not possible to achieve equidistant time intervals since it was considered to be more important to use fixed survey days every week in order to make it easier for participants to remember the survey schedule and therefore to increase participation rates. The problem of unequal time intervals can be addressed, for example, by a simple procedure (see Appendix A in Asparouhov et al. (2018), Kelava \& Brandt, 2019) using phantom variables (Rindskopf, 1984). As an incentive to participate, there was a weekly prize draw for all those students who participated in all surveys of that week. Third, performance and attendance information for the accompanying tutorial sessions was collected once a week. One of the reasons for collecting data weekly rather than waiting to collate all this information at the end of the semester was to be able to obtain information on drop outs during the semester.

Participation in each part of the study was voluntary and students' written consent was obtained for the use of the information they had given for the purposes of the study and for matching the different data sources via anonymous codes. The sample size given above refers to the number of students who had given consent to participate in all three types of data collection (including participation in at least one online survey).

\subsection{Sample properties}
Our sample consists of 122 students with an average (median) age of 19.60 (19) ($SD = 1.49$). 55 (45.08\%) students indicated to be female and 66 (54.09\%) students to be male. 
The sample includes 14 students who major in mathematics B.Sc. (3 females), 57 teacher candidates in mathematics B.Ed. (39 females) and 44 in physics B.Sc. (9 females). A further 6 students were voluntary participants enrolled on other study programs. The sample mean for German GPA (Abitur) was $1.86$ ($SD = 0.53$) (with 1 being the best grade and 4 the lowest pass grade) and the mean for the final math grade from school was $12.03$ ($SD = 2.30$) (with 15 being the highest possible grade and 5 the lowest pass grade). No significant differences in these measures were found between female and male students.

\subsection{Measures}
In the study, a broad range of scales was used to address the wide variety of variables related to student drop out (e.g., Bean, 20005; Burrus et al., 2013).

\subsubsection{Initial assessment.} The initial assessment included information on stable traits as well as initial states of the students. 
A collection of approved TIMSS items (e.g., Mullis, Martin, Ruddock, O'Sullivan, Arora, \& Erberber, 2005) was used for the assessment of students' a priori math performance. The mean was $10.53$ ($SD = 3.51$) out of 20 possible points with significant differences between female ($9.13$, $SD = 3.09$) and male ($11.65$, $SD = 3.44$) students ($p < .001$). In order to obtain further information on students' cognitive abilities (IQ), we used a German adaption of the Culture Fair Intelligence Test Scale 3 (CFT-3, Cattell \& Weiß 1980) (mean = $116.93$, $SD = 12.16$). As expected, the math students in our sample showed higher cognitive abilities ($t = 15.37$, $p < .001$) than the average of the corresponding population (with mean = 100 (SD = 15) of 17-18 year-old students with similar educational backgrounds).

In addition to these performance measures, information was available on students’ educational backgrounds such as high school performance, type of school attended and whether they had attended preparatory courses before university or an additional in-depth mathematics class at school. Socio-demographic measures included parental educational qualifications, whether students were from immigrant families, and their financial situation. Professional interests were measured using the AIST-R questionnaire which is based on Holland's (1997) RIASEC model (Bergmann, \& Eder, 2005). We also obtained information on motivational aspects (Wigfield \& Cambria, 2010), personality (Big Five personality traits - BFI-2-XS; Soto \& John, 2017)), and on positive and negative affect (using the positive and negative affect schedule PANAS; Watson, Clark, \& Tellegen, 1988). Locus of control was measured through the IE-4 scale (Kovaleva, Beierlein, Kemper, \& Rammstedt, 2012).

\subsubsection{State measures.} After this initial assessment, information on intra-individual changes was collected via the online surveys as described above. The items re-assessed students' motivational states and the positive and negative affective states which had first been assessed via the initial questionnaire. In addition, data were collected on students' current intention to quit their course, fear of failure, subjective feeling of being overwhelmed by the demands of the course, ability to follow the calculus course, assignments, and learning behavior. The items can be obtained from the authors on request.

In sum, the data can be characterized as multivariate observations on several days a week, integrating within-person and across-person variability as well as objective and subjective data.


\subsection{Model}
In this subsection, we describe the specified model that we used for the analysis of the ILD and the forecasting. 
The aim of the model is to identify and to forecast which student may drop out and when she would do so. For this, we need information about dynamic intra-individual changes (on the within-level) as well as inter-individual differences (on the between-level). In practical terms, we would like to describe what kind of responses will likely result in future drop out (e.g., being afraid to fail, PAN, and low cognitive skills). This implies (a) we have a (between-level) risk factor at the beginning of the math study (e.g., cognitive skills), and (b) if persons express strong (within-level) affective states, tailored interventions can be imposed in a timely fashion.

Note that, in the following, we will assume that the observed variables are continuous (and conditionally normal; see below). However, the general NDLC-SEM framework can handle also discrete (categorical, ordered categorical, count) data (for details see Kelava \& Brandt, 2019).

\paragraph{Within-level measurement model.}
On the within level, 17 observed variables were used to operationalize seven latent factors. All observed variables were centered using the mean from the first time point and inverted if necessary such that we expected higher values for persons with an intention to quit. We used the following state variables: Three observed variables measured the \textit{subjective importance of the content} (i.e., attainment value in motivation theory; e.g., ``content not important'') and two variables indicated how much of their time students felt they needed to invest in order to understand the content at the expense of other important activities (i.e., \textit{cost} in motivation theory; e.g., ``too much time''). Two observed variables were directly related to the \textit{intention to quit} and measured thoughts about quitting or fears to fail the math course (e.g., ``afraid to fail''). Two observed variables (e.g., ``no understanding'') measured subjective assessment of students' own \textit{understanding}. Two observed variables (e.g., ``\textit{stress}'') measured how stressed students felt. Finally, three observed variables each (``no PAP'' and ``PAN'') measured the \textit{positive} and \textit{negative affective states}. We assumed that measurement models were the same whether students intended to quit or not:
\begin{align}
[Y_{1it} | S_{it} = s] & = \Lambda_{10} \eta_{1its} +
\epsilon_{1it}, 
\label{eqex:l1:m1}
\end{align}
with class-invariant normally distributed residual terms $\epsilon_{1ijt}\sim \mathrm{N}(0,\sigma_{\epsilon_{1j}}^2)$ ($j=1\ldots 17$). $\eta_{1its}$ was the ($7 \times 1$) vector of the seven normally distributed latent variables (see below). The ($17 \times 7$) factor loading matrix $\Lambda_{10}$ used a simple structure that represented the factor structure described above and standard identification constraints. Note that, in Eq. \eqref{eqex:l1:m1}, $\eta_{1its}$ describes individual-specific system-dynamics (e.g., intra-individual changes of affective states), which will be further explained. $\epsilon_{1it}$ contains person- and time-specific residuals that are unexplained by the actual process.

\paragraph{Between-level measurement model.}
On the between level, we specified one latent construct $\eta_2$ (``students' cognitive abilities''; IQ) based on three centered items on cognitive tasks (CFT-3) that were measured at baseline:
\begin{align}
[Y_{2i}] & = \Lambda_{2} \eta_{2i} + \epsilon_{2i}, 
\label{eqex:l2:m1}
\end{align}
with a ($3 \times 1$) factor loading matrix $\Lambda_{2}$, a normally distributed latent factor $\eta_{2i}\sim \mathrm{N}(0,\sigma_{\zeta_{3}}^{2})$, and  normally distributed residual terms $\epsilon_{2ij}\sim \mathrm{N}(0,\sigma_{\epsilon_{2j}}^2)$ ($j=1\ldots 3$) .

\paragraph{Within-level structural model.}
We specified an AR(1) model for the seven continuous latent state variables using a class-specific nonlinear model that accounted for the hypothesis that cognitive pre-university measures moderate the motivational and self-regulatory aspects during university. \footnote{Note that we tested different ARIMA-type models. The final model was an AR(1) model.}
We assumed that persons with an intention to quit generally have higher scores in the seven factors and that the relationships between the variables can change across the intention states (classes):
\begin{align}
[\eta_{1it} | S_{it} = s] & = 
\alpha_{1is} + 
B_{1is} \eta_{1i(t-1)} +
\zeta_{1it}, 
\label{eqex:l1:s1}
\end{align}
where the ($7 \times 1$) vector of intercepts $\alpha_{1is}$ and the ($7 \times 7$)  matrix $B_{1is}$ were modeled as random effects on the between level (see below). For the AR(1) structure, we specified $B_{1is}$ as a diagonal matrix, and normally distributed class-invariant residual terms $\zeta_{1ijt}\sim \mathrm{N}(0,\sigma_{\zeta_{1j}}^2)$ ($j=1\ldots 7$). 

\paragraph{Between-level structural model.}

For the random intercept vector $\alpha_{1is}$ specified on the within level and the AR coefficient matrix, the following two models were used:
\begin{align}
\alpha_{1is} &= \alpha_{21s} + B_{2} \eta_{2i} + \zeta_{2i} \label{eqex:l2:alpha1}\\
B_{1is} & = B_{1s} + \Omega_{2s} \eta_{2i}
\label{eqex:l2:B1}
\end{align}
where $\alpha_{2s1}$ is a ($7 \times 1$) class-specific intercept vector, $B_{2}$ is the ($7 \times 1$) vector specifying the main effect of IQ on the within-level variables. $B_{1s}$ is the ($7 \times 7$) diagonal matrix of auto-regressive coefficients and the ($7 \times 7$) diagonal matrix $\Omega_{2}$ includes the cross-level interactions that indicate whether IQ moderates the within-level relationships. $\zeta_{2i}$ included all remaining unexplained stable inter-individual differences.

\paragraph{Markov switching model.}
The Markov switching model for persons who were still in the no-intention state was set up as 
\begin{align}
P(S_{it} = 1| S_{i(t-1)} = 1) & = \frac{\exp(\nu_{it11})}{\Sigma_{k=1}^{K}\exp(\nu_{itk1})}\label{eqex:markov:NDLC-SEM1}
\end{align}
The probability to stay in state $S=1$ at $t$, i.e. no intention to quit, with no intention to quit at $t-1$ was modeled via
\begin{align}
\nu_{it11}&=\gamma_{1} + \gamma_{2} \eta_{2i}  + \Gamma_{3} \eta_{1i(t-1)} + \Gamma_{4} \eta_{1i(t-1)}\eta_{2i}
\label{eqex:markov:NDLC-SEM2}
\end{align}
where $\gamma_{1}$ is an intercept, $\gamma_{2}$ included the effects of the baseline covariate IQ ($\eta_{2i}$), the ($ 1 \times 7$) coefficient vector $\Gamma_{3}$ the lagged effects of the within-level variables, and the ($ 1 \times 7$) coefficient vector $\Gamma_{4}$ included three interaction effects between IQ and the self-regulatory variables from the within level (too much time, afraid to fail, stress). 

The probability to return to state $S=1$ at $t$, after an intention to quit at $t-1$ was modeled via a uniform prior $P(S_{it} = 1| S_{i(t-1)} = 2)\sim unif(.0, .1)$. This informative prior was used in order to be able to better interpret the states.

Finally, the state variables were predicted using a binomial distribution: 
\begin{align}
S_{it}&\sim Bin(P(S_{it} = s_k | S_{i(t-1)} = s_l))\quad s_k, s_l =1,2
\label{eqex:markov:NDLC-SEM3}
\end{align}
Note that if a manifest drop-out for student $i$ occurred at time point $t$, $S_{it}=\ldots =S_{iT}=2$ become observed values. Before that time point, $S$ is a discrete latent state (class) variable and no direct information is available. This ensures identification of the latent discrete state variable that is in line with the nomenclature ``intention to quit''.

\begin{figure}
\begin{center}
\includegraphics[width=.75\textwidth]{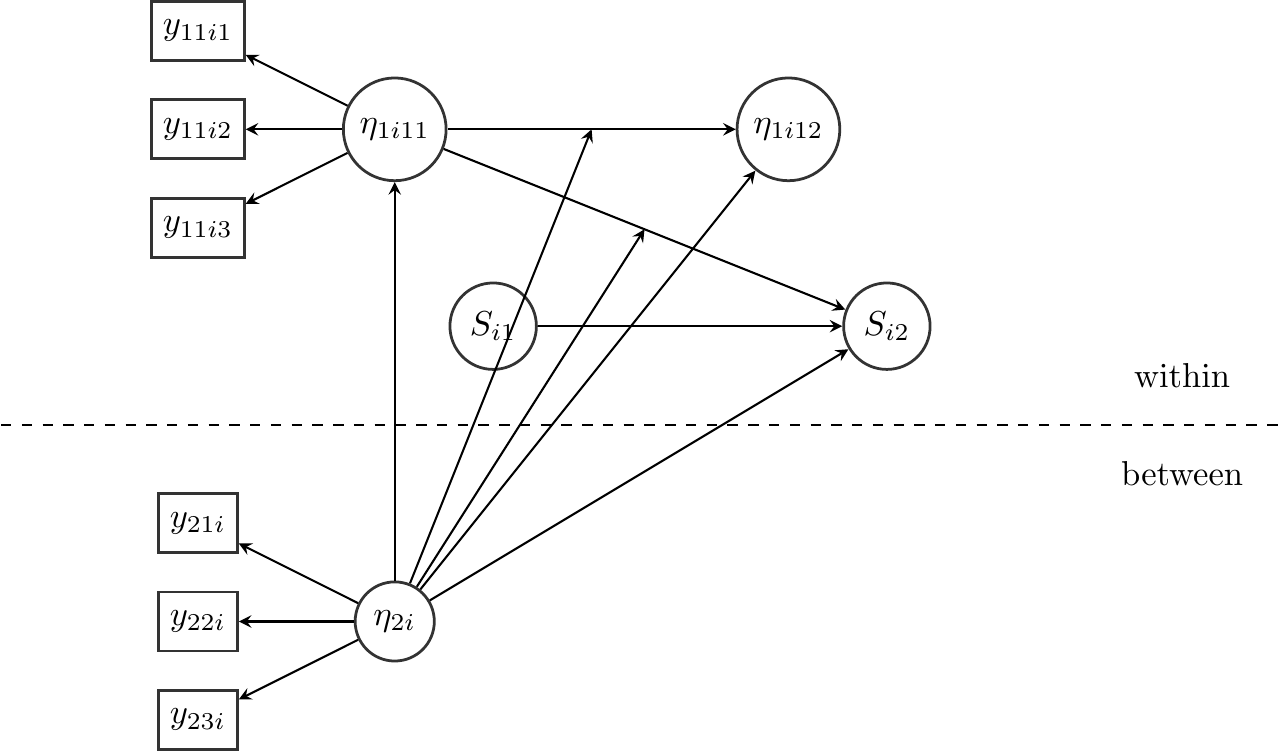}
\end{center}
\caption{Path diagram for the final model for the seven time-series for the first two measurement occasions (out of 50). There are seven within-level scales.}
\label{fig:path1}
\end{figure}

The complete model is depicted in Figure \ref{fig:path1} for the first two measurement occasions. 

\paragraph{Priors for the baseline model.}
The prior distributions are summarized in the following equations \footnote{Note that $j$ is a count index. It is used to index elements within a vector. Its length (e.g., $j=1...3$ or $j=1...7$) is redefined in every equation. The count index is used to save space.}:

\begin{align}
\lambda_{10j} & \sim \mathrm{N}(0,1)^+  &j=1\ldots 6\label{eq:prioralpha}\\
\lambda_{2j}  & \sim \mathrm{N}(0,1)^+  &j=1\ldots 2\\ 
\alpha_{21(s=1)j} & \sim \mathrm{N}(0,1)  &j=1\ldots 7\\ 
\Delta \alpha_{21(s=2)j} &\sim \mathrm{N}(0,1)^+,\quad \alpha_{21(s=2)j} = \alpha_{21(s=1)j}+\Delta \alpha_{21(s=2)j} &j=1\ldots 7\\ 
\beta_{2(s=1)j}   & \sim \mathrm{N}(0,1)  &j=1\ldots 7\\ 
\Delta \beta_{2(s=2)j} &\sim \mathrm{N}(0,1),\quad \beta_{2(s=2)j} = \beta_{2(s=1)j}+\Delta \beta_{2(s=2)j} &j=1\ldots 7\\ 
\beta_{1(s=1)j}  & \sim \mathrm{N}(0,1)  &j=1\ldots 7\\ 
\Delta \beta_{1(s=2)j} &\sim \mathrm{N}(0,1),\quad \beta_{1(s=2)j} = \beta_{1(s=1)j}+\Delta \beta_{1(s=2)j} &j=1\ldots 7\\
\omega_{2(s=1)j}  & \sim \mathrm{N}(0,1)  &j=1\ldots 7\\ 
\Delta \omega_{2(s=2)j} &\sim \mathrm{N}(0,1),\quad \omega_{2(s=2)j} = \omega_{2(s=1)j} + \Delta \omega_{2(s=2)j} &j=1\ldots 7\\ 
\gamma_{1}   & \sim \mathrm{N}(0,1)  &\\
\gamma_{2}   & \sim \mathrm{N}(0,1)  &\\
\gamma_{3j}   & \sim \mathrm{N}(0,1)  &j=1\ldots 7\\
\gamma_{4j}   & \sim \mathrm{N}(0,1)  &j=1\ldots 3\\
\sigma_{\zeta_{1j}}^{-2} & \sim Gamma(9,4)&j=1\ldots 7\\
\sigma_{\zeta_{2j}}^{-2} & \sim Gamma(9,4)&j=1\ldots 7\\
\sigma_{\zeta_{3}}^{-2} & \sim Gamma(9,4)\\
\sigma_{\epsilon_{1j}}^{-2}& \sim Gamma(9,4) &j=1\ldots 17\\
\sigma_{\epsilon_{2j}}^{-2}& \sim Gamma(9,4) &j=1\ldots 3\label{eq:prioromega}
\end{align}
where $Gamma(a,b)$ is the Gamma distribution with hyperparameters $a,b$. Note that $\Delta \alpha_{21(s=2)j}$ is a censored distribution, that is, we are assuming that persons who switch to $S=2$ (intention to quit) have higher values in the overall level of the factor. All items were inverted if necessary to ensure a straightforward interpretation.

\paragraph{Identification}
The identification of the model relates to two major aspects. First, the (continuous) latent variables (factors) need to be identified; this is achieved by using standard constraints for structural equation models (e.g., by use of a scaling item with factor loading fixed to one). Second, the latent discrete states need to be identified. This identification includes more aspects and is driven by actual differences in the observed response patterns. In addition, this identification is strongly related to the interpretation of the discrete state as ``intention to drop out''. In order to achieve both, we follow a similar strategy as Jeon (2019). She describes how mixtures can be extracted using a more confirmatory method instead of traditional exploratory methods (such as Bauer, 2005). First, we use a censored distribution for the scale means such that persons in the state S=2 are constrained to have higher scores in all seven negative affect scales -- which is in line with expectations about drop-out. Second, we use these scales and theory-derived interactions with cognitive skills to predict the probabilities to switch from one latent discrete state to the other. Third, we restrict the probability to switch back, that is, we ensure that persons who have a high probability to quit, do not change their mind at the following time points -- again this is an assumption about the interpretation of the discrete state that is imposed using this confirmatory approach. And fourth, we use a partially observed latent class variable because information about persons who actually dropped out were available. 

\paragraph{Implementation}
The model was implemented in Jags 4.2 (Plummer, 2017) and run via the R2jags package. Four chains each with 30,000 iterations were run with 25,000 iterations burn-in. Convergence was checked graphically via trace plots and density plots. The Rhat statistic was used to assess chain mixing and a criterion of Rhat$<1.12$ was achieved for all parameters. Computation was conducted on a virtual machine of a server using 4 cores of an Intel XEON Gold 6244 3.6GHz Processor each with and 40GB RAM; computation time for the empirical example was 23 hours. The computational burden for the simulation study was considerably lower for each replication, as only 3 latent within-level factors were used and sample sizes were smaller; for example, the condition with $N_1=50,N_t=25$ took about 50 minutes for each replication on the same server.


\section{Forecast implementation}
In this section, we describe a \textit{Forward Filtering Backward Sampling (FFBS)} method which we will use in our empirical example and simulation study. It is an adaptation of previous work by West and Harrison (1997). In order to make use of their work, we need to reformulate our model in terms of observation and system equations. The observation equation relates the observed entity to latent continuous states; the system equation describes the evolution of these latent continuous states over time. Here, we suggest forecasting the within-level latent factors $\eta_{1ijt}$. For the $N_1$ dimensional vector $\eta_{1jt}=(\eta_{11jt},\ldots ,\eta_{1N_1jt})'$ for the $j$-th factor, and our model description above, these equations are within each state $S_t=s$:
\begin{align}
\eta_{1jt}&=\mathbf{F}_{jt}\boldsymbol\theta_{jts}+\mathbf{v}_{jts},\quad \mathbf{v}_{jts}\sim N(\mathbf{0},\mathbf{V}_{jts})\\
\boldsymbol\theta_{jts}&=\mathbf{G}_{jts}\boldsymbol\theta_{j,t-1,s}+\mathbf{w}_{jts},\quad \mathbf{w}_{jts}\sim N(0,\mathbf{W}_{jts})
\end{align}
with $\mathbf{F}_{jt}=(\mathbf{1}_{N_1},\mathbf{0}_{N_1},\zeta_{2j},\eta_{2j},\eta_{2j}\cdot \eta_{1j,t-1})$, an identity matrix $\mathbf{G}_{jts}=\mathbf{I}_5$, $\mathbf{V}_{jts}=
\mathbf{I}_{N_1}\cdot\sigma_{\zeta_1}^2$, and $\mathbf{W}_{jts}=\mathbf{0}_{5\times 5}$. In general, $\mathbf{F}$ includes all known values of independent variables; examples for the specification of $\mathbf{F}$ (e.g., specific auto-regressive models or regression models) can be found in  West and Harrison (1997). Here, $\boldsymbol\theta_{jts}$ is a $5$-dimensional system vector and $\mathbf{v}_{jts}$ is a so-called observational error. Since our model does not include time-depending changes in the actual parameters (such as $B_{1s}$), the errors of the system equation are zero. Often, these equations are summarized as a quadruple $\{\mathbf{F}_{jt},\mathbf{G}_{jts}, \mathbf{V}_{jts}, \mathbf{W}_{jts} \}$, which simplifies in our example to $\{\mathbf{F}_{jt},\mathbf{G}, \mathbf{V}, \mathbf{W}\}$ (because these matrices do not change over time) that holds with probability $\pi(S_t)$ within each state (see next subsection).

Note that in this formulation, random effects ($\zeta_{2j}$) are not marginalized out; instead a person-specific estimate of the forecast is conducted using the sampled scores for $\zeta_{2j}$ from the posterior distribution. An alternative formulation that provides forecasts with marginalized random effects can be found in Gamerman and Migon (1993). The main difference between these two methods is that the formulation here allows us to forecast at an individual level as expressed with ($\zeta_{2j}$), whereas the formulation in Gamerman and Migon (1993) takes the variance of the random effects ($\sigma_{\zeta_{2j}}^{2}$) into account for the precision of the forecast intervals. 
Extensions to forecasts of the observed variables (i.e., the individual observed item responses $Y_{1it}$) is straightforward and can be found, for example, in chapter 3.3.7 in Petris, Petrone, and Campagnoli (2009).

\subsection{Model probability at time $t$}
At time $t$, corresponding to the two latent discrete states, there are the following two models (cf. West \& Harrison, 1997, Chapter 12):
\begin{align}
    M_{t}(S_t): & \quad \{\mathbf{F}_{jt},\mathbf{G}, \mathbf{V}, \mathbf{W} \}
\end{align}

We assume that for each time $t$, a model $M_{t}(S_t)$ applies with a probability:
\begin{align}
    \pi(S_t) & = Pr[M_t(S_t) | D_{t-1}]
\end{align}
dependent on previous data $D_{t-1}$.

\subsection{Prior for the state variable at $t=0$}
At time $t=0$, the prior for the state variable is given as:
\begin{align}
    (\boldsymbol{\theta}_{t=0} | D_0) & \sim \mathrm{N}[\textbf{m}_0, \textbf{C}_0]
\end{align}
where the expectation $\textbf{m}_0$ and covariance matrix $\textbf{C}_0$ are assumed to be known at $t=0$.

\subsection{Probability of the state variable at time $t-1$}
Previous data $D_{t-1}$ is summarized in terms of a two-component mixture distribution for $\boldsymbol{\theta}_{t-1}$ which is based on the two possible models $M_{t-1}(S_{t-1})$ (depending on $S_{t-1}$) obtained at time $t-1$.
For each $S_{t-1} = 1,2$, the model $M_{t-1}(S_{t-1})$ has posterior probability $p_{t-1}(t-1)$ which is fixed and known at $t-1$.
Given the model $M_{t-1}(S_{t-1})$ and data $D_{t-1}$, each mixture component of $\boldsymbol{\theta}_{t-1}$ has the following probability:
\begin{align}
    ( \boldsymbol{\theta}_{t-1} | M_{t-1}(S_{t-1}), D_{t-1}) & \sim \mathrm{N}[\textbf{m}_{t-1}(S_{t-1}), \textbf{C}_{t-1}(S_{t-1})]
\end{align}

Note that $\textbf{m}_{t-1}(S_{t-1})$ and $\textbf{C}_{t-1}(S_{t-1})$ depend on the model applying at $t-1$ (see also Eq. \eqref{eq:theta_t}).

\subsection{Probability of the state variable at time $t$}
At time $t$, $\boldsymbol{\theta}_{t}$ and $\eta_{1jt}$ depend on the combinations of possible models at both $t-1$ and $t$.
For each combination of $S_{t-1}$ and $S_t$, we have
\begin{align}
    (\boldsymbol{\theta}_{t} | M_{t}(S_t), M_{t-1}(S_{t-1}), D_{t-1}) & \sim \mathrm{N}[\textbf{a}_t(S_{t-1}),\textbf{R}_t(S_t, S_{t-1})] \quad \text{with}
    \label{eq:theta_t}\\
    \textbf{a}_t(S_{t-1}) & = \textbf{G} \textbf{m}_{t-1}(S_{t-1})\\
    \textbf{R}_t(S_t, S_{t-1}) & = \textbf{G} \textbf{C}_{t-1}(S_{t-1}) \textbf{G}' + \textbf{W}
\end{align}

\subsection{One-step ahead forecast}
The one-step ahead forecast distribution is given, for each of the four possible combination of models, by:
\begin{align}
    (\eta_{1jt} | M_{t}(S_t), M_{t-1}(S_{t-1}), D_{t-1}) & \sim \mathrm{N}[f_t(S_{t-1}), Q_t(S_t,S_{t-1})]  \quad \text{with}\\
    f_t(S_{t-1}) & = \textbf{F}_{jt}' \textbf{a}_t(S_{t-1})\\
    Q_t(S_t,S_{t-1})& = \textbf{F}_{jt}' \textbf{R}_t(S_t,S_{t-1})\textbf{F}_{jt} + \textbf{V}
\end{align}

\subsection{Probability of each model combination}
Obtaining the forecast distribution unconditional on possible models involves mixing the normal components utilizing their probabilities. For each combination of $S_t$ and $S_{t-1}$, the probabilities are given as:
\begin{align}
    \mathrm{Pr}[M_{t}(S_t), M_{t-1}(S_{t-1}) | D_{t-1}] & = \mathrm{Pr}[M_{t}(S_t) | M_{t-1}(S_{t-1}), D_{t-1}] \,  \mathrm{Pr}[M_{t-1}(S_{t-1}) | D_{t-1}]
\end{align}

By the assumption from above, models apply at $t$ with probabilities $\pi(S_t)$:
\begin{align}
    \mathrm{Pr}[M_{t}(S_t) | M_{t-1}(S_{t-1}), D_{t-1}] & = \pi(S_t)\\
    \mathrm{Pr}[M_{t-1}(S_{t-1}) | D_{t-1}] &= p_{t-1}(S_{t-1})
\end{align}
such that
\begin{align}
    \mathrm{Pr}[M_{t}(S_t), M_{t-1}(S_{t-1}) | D_{t-1}] & = \pi(S_t) p_{t-1}(S_{t-1})
\end{align}

Note that this step is used to forecast the discrete latent time-dependent states (e.g., intention to quit).

\subsection{Marginal predictive density}
Using these probabilities, the marginal predictive density for $\eta_{1jt}$ is the mixture of the corresponding four components:
\begin{align}
    p(\eta_{1jt} | D_{t-1}) & = \sum_{S_t=1}^{2} \sum_{S_{t-1}=1}^{2} \{ \pi(S_t) p_{t-1}(S_{t-1}) p(\eta_{1jt} | M_{t}(S_t), M_{t-1}(S_{t-1}), D_{t-1})\}
\end{align}

Note that this step is used to forecast the continuous latent within-factor $\eta_{1jt}$ (e.g., an affective state).

\subsection{Updating the prior distributions}
When $\eta_{1jt}$ is realized (i.e. via sampling from the posterior), the prior distributions are updated for given $S_{t-1}$ and $S_t$ (for details see West \& Harrison, 1997).
\begin{align}
    (\boldsymbol{\theta}_{t} | M_{t}(S_t), M_{t-1}(S_{t-1}), D_t) & \sim \mathrm{N}[\textbf{m}_t(S_t, S_{t-1}),\textbf{C}_t(S_t, S_{t-1})] \quad \text{with}\\
    \textbf{m}_t(S_t, S_{t-1}) &= \textbf{a}_t(S_{t-1}) + \textbf{A}_t(S_t,S_{t-1})e_t(S_{t-1})\\
    \textbf{C}_t(S_t, S_{t-1}) &= \textbf{R}_t(S_t,S_{t-1}) - \textbf{A}_t(S_t,S_{t-1}) \textbf{A}_t(S_t,S_{t-1})' Q_t(S_t,S_{t-1})\\
    e_t(S_{t-1}) & = \eta_{1jt} - f_t(S_{t-1})\\
    \textbf{A}_t(S_t,S_{t-1}) & = \textbf{R}_t(S_t,S_{t-1}) \textbf{F}_{jt} / Q_t(S_t,S_{t-1})
\end{align}

\subsection{Posterior probabilities}
The posterior probabilities across the four possible models are given as:
\begin{align}
    p_t(S_t,S_{t-1}) & = \mathrm{Pr}[M_{t}(S_t), M_{t-1}(S_{t-1}) | D_{t}] \\
    & \propto \pi(S_t) p_{t-1}(S_{t-1}) p(\eta_{1jt} | M_{t}(S_t), M_{t-1}(S_{t-1}), D_{t-1})
\end{align}
with the predictive density 
\begin{align}
    p_t(S_t, S_{t-1}) & = \frac{c_t \pi(S_t) p_{t-1}(S_{t-1})}{Q_t(S_t,S_{t-1})^{1/2}} \exp\{-.5 e_t(S_t,S_{t-1})^2 / Q_t(S_t,S_{t-1})\}
\end{align}
with the normalization constant $c_t$ such that:
\begin{align}
    \sum_{S_t=1}^{2} \sum_{S_{t-1}=1}^{2} p_t(S_t,S_{t-1}) = 1
\end{align}

\subsection{Average posterior model probabilities}
The posterior model probabilities from above average using the four 
component mixtures, such that inferences about $\boldsymbol{\theta}_{t}$ are possible:
\begin{align}
    p(\boldsymbol{\theta}_{t} | D_t) & = \sum_{S_t=1}^{2} \sum_{S_{t-1}=1}^{2} p(\boldsymbol{\theta}_{t} | M_{t}(S_t), M_{t-1}(S_{t-1}), D_{t})
    p_t(S_t,S_{t-1})
\end{align}

These calculations complete the evolution and updating steps at time $t$. The last equation can also be expressed as:
\begin{align}
    p(\boldsymbol{\theta}_{t} | D_t) & = \sum_{S_t=1}^{2} p(\boldsymbol{\theta}_{t} | M_t(S_t), D_t) p_t(S_t) \quad \text{with}\\
    p(\boldsymbol{\theta}_{t} | M_t(S_t), D_t) & = \sum_{S_{t-1}=1}^{2} p(\boldsymbol{\theta}_{t} | M_{t}(S_t), M_{t-1}(S_{t-1}), D_t) p_t(S_t, S_{t-1})/p_t(S_t)
\end{align}
or as an approximation:
\begin{align}
    (\boldsymbol{\theta}_{t} | M_t(S_t), D_t) & \sim \mathrm{N}[\textbf{m}_t(S_t),\textbf{C}_t(S_t)] \quad \text{with}\\
    \textbf{m}_t(S_t) & = \sum_{S_{t-1}=1}^{2} \textbf{m}_t(S_t,S_{t-1}) p_t(S_t,S_{t-1})/p_t(S_t)\\
    \textbf{C}_t(S_t) & = \sum_{S_{t-1}=1}^{2} \{ \textbf{C}_t(S_t,S_{t-1}) + (\textbf{m}_t(S_t) - \textbf{m}_t(S_t,S_{t-1})) (\textbf{m}_t(S_t) - \textbf{m}_t(S_t,S_{t-1}))'\} \times \nonumber \\
    & \quad p_t(S_t,S_{t-1})/p_t(S_t)
\end{align}

This concludes the Forward Filtering Backward Sampling (FFBS) algorithm. Next, we will use this procedure for the forecasting with our real empirical data.

\section{Results for the empirical example}
In this section, we present the results for the empirical example predicting student drop out from math. First, we will describe the results for the parameter estimates. Then we will provide information about the forecast.

\subsection{Overall model results}

\begin{figure}
	\centering
	\includegraphics[angle=0,width=1\textwidth,page=1]{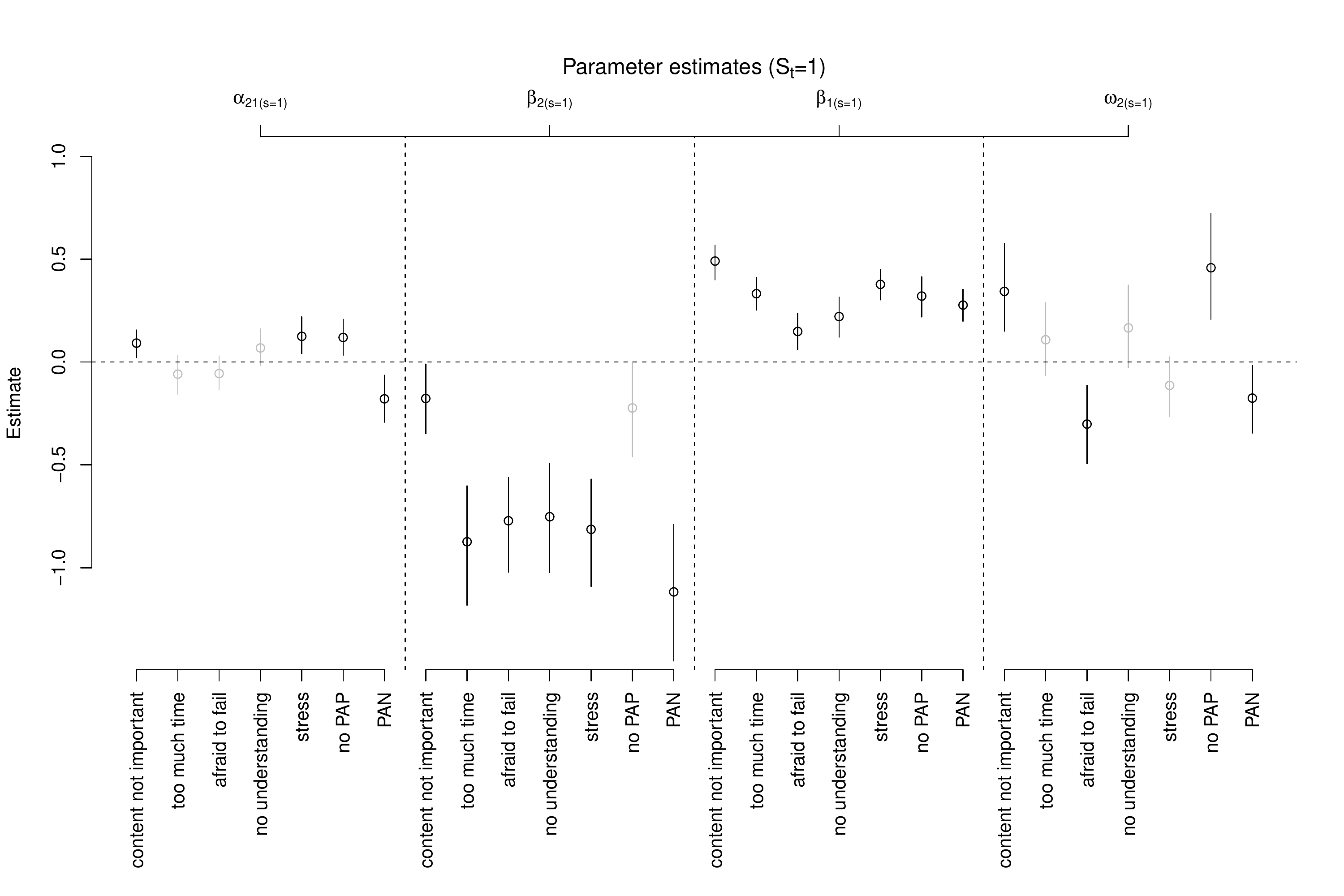}
	\caption{State-specific parameter estimates from the empirical example for the discrete state $S_t=1$ (no intention to quit). Significant estimates (i.e., the 95\% credible interval did not include zero) are indicated in black.}
	\label{fig:pars1}
\end{figure}

\begin{figure}
	\centering
	\includegraphics[angle=0,width=1\textwidth,page=2]{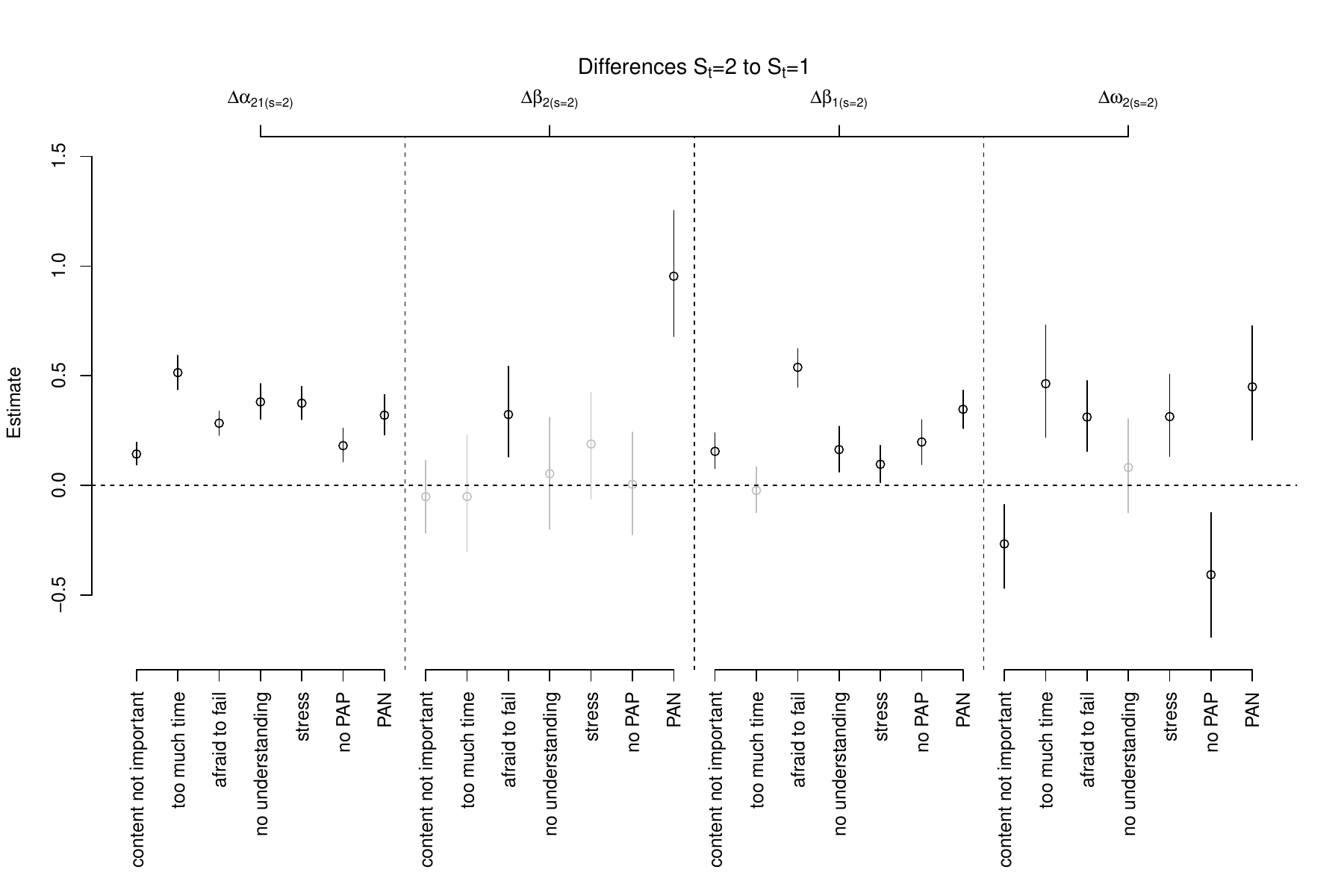}
	\caption{State-specific parameter estimates from the empirical example for the discrete state $S_t=2$ (intention to quit). Significant differences between discrete states $S_t=1$ and $S_t=2$ (i.e., the 95\% credible interval did not include zero) are indicated in black.}
	\label{fig:pars2}
\end{figure}

\begin{figure}
	\centering
	\includegraphics[angle=0,width=1\textwidth,page=4]{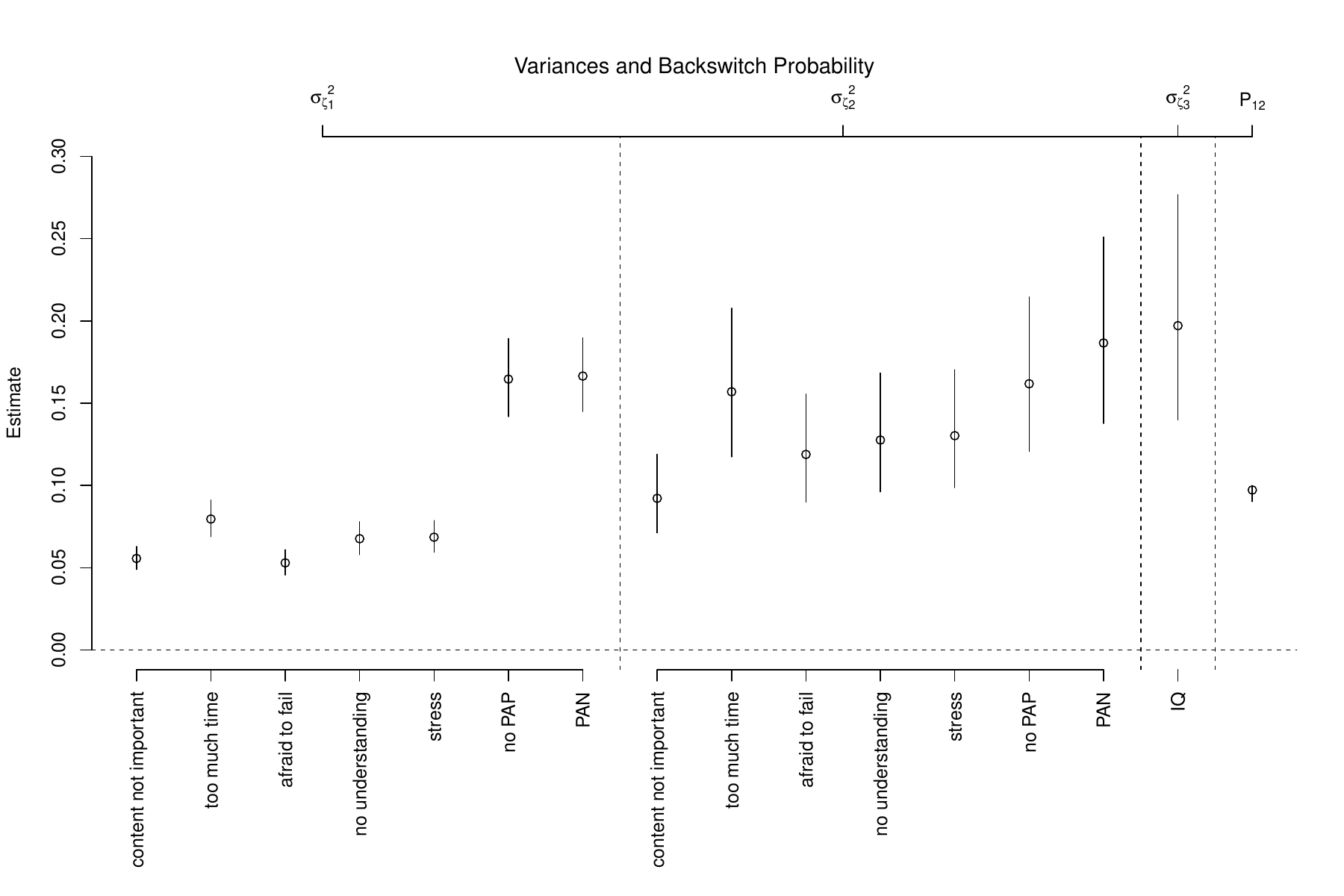}
	\caption{Parameter estimates from the empirical example for the variances at within and between levels as well as backswitch probability ($P_{12}$). Significant estimates (i.e., the 95\% credible interval did not include zero) are indicated in black.}
	\label{fig:pars3}
\end{figure}

\begin{figure}
	\centering
	\includegraphics[angle=0,width=1\textwidth,page=3]{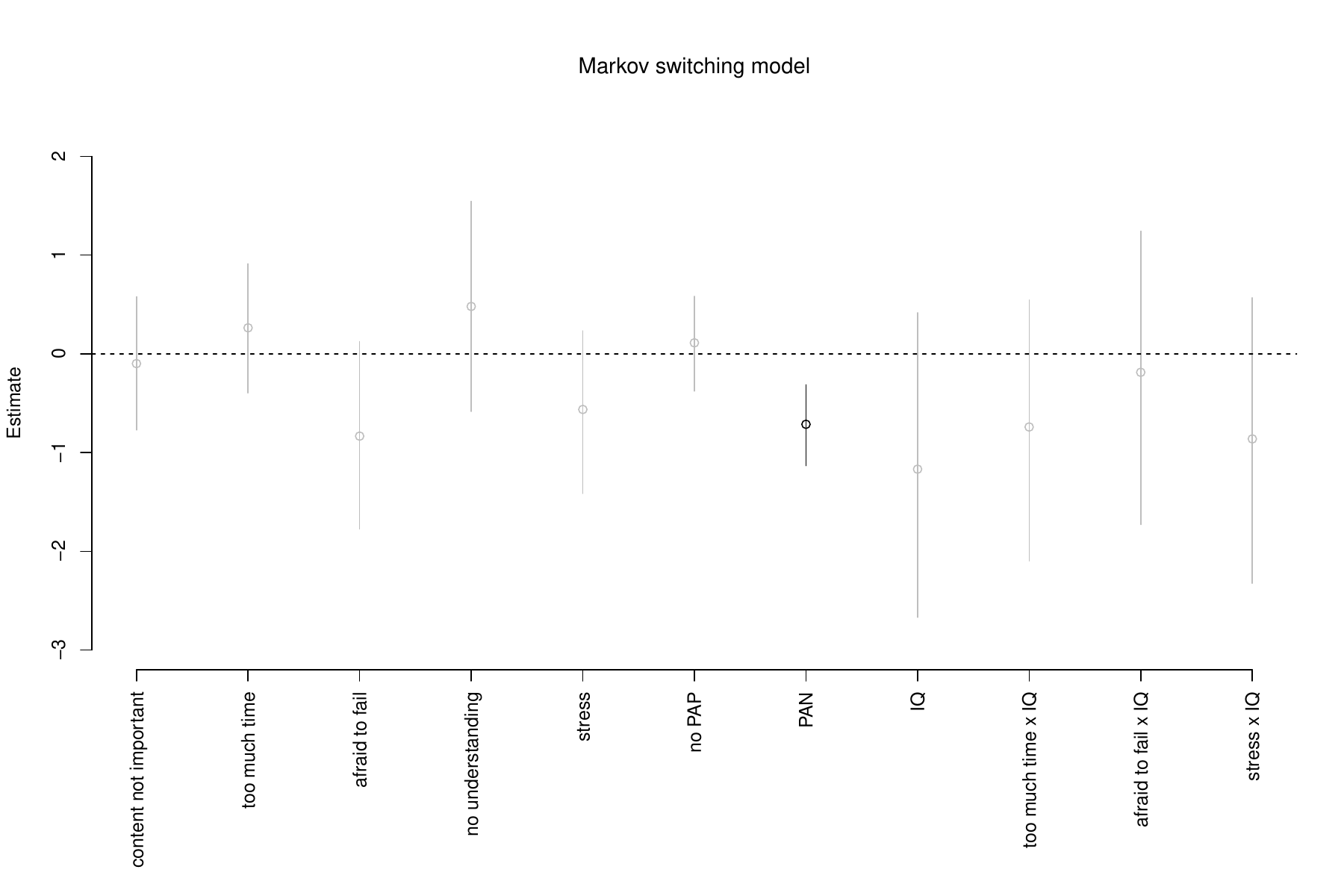}
	\caption{Parameter estimates from the empirical example for the Markov Switching Model. Significant estimates (i.e., the 95\% credible interval did not include zero) are indicated in black. Negative estimates indicate that persons were more likely to switch to the discrete state $S_t=2$ (intention to quit).}
	\label{fig:pars4}
\end{figure}

Parameter estimates and credible intervals for all parameters are depicted in Figures~\ref{fig:pars1} to \ref{fig:pars4}. 

Figure~\ref{fig:pars1} shows the results for parameters on the latent discrete state $S_t=1$ (no intention to quit). 
The baseline scale for \textit{cognitive skills} (IQ; indicated by $\beta_{2(s=1)}$) was predictive for all seven within-scales except for the \textit{positive affect} (\textit{no PAP}). Parameter estimates for this construct were negative, that is, persons with higher cognitive skills had lower scores on the within level scales such as \textit{stress}. 
The auto-regressive coefficients ($\beta_{1(s=1)}$) were positive for all seven within-scales under (no intention to quit). This implied that students' responses on these scales followed a regular pattern.  
Several \textit{interactions between the within-scales and the cognitive skills} could be observed (indicated with $\omega_{2(s=1)}$): For \textit{being afraid to fail} and \textit{negative affect (PAN)} they were negative under $S_t=1$, implying that higher \textit{cognitive skills} could diminish the negative effects of these scales. For the scales \textit{content not important} and \textit{no PAP} the interaction effects were positive, implying that persons who simultaneously had higher \textit{cognitive skills} and expressed that they did not find the content important were even more stable on this expression over time. 

Figure~\ref{fig:pars2} depicts the differences in the parameters between discrete states $S_t=1$ and $S_t=2$ (intention to quit). 
Persons who switched from the discrete state $S_t=1$ (no intention to quit) to $S_t=2$ (intention to quit) showed consistently higher values on all seven scales ($\Delta\alpha_{21(s=2)}$). This provided evidence that the latent states actually indicated attitudes that can be considered to be related to an intention to quit.
The effects of the cognitive skills were similar in state $S_t=2$ (intention to quit) compared to $S_t=1$ (indicated by $\Delta\beta_{2(s=2)}$) except for \textit{PAN} and \textit{being afraid to fail}. For these two scales, the effect of cognitive skills was smaller, that is, persons with different cognitive skills had similar scores on these two scales (e.g. for \textit{PAN} $\beta_{2(s=1)7}=-1.117$ under $S_t=1$ vs. $\beta_{2(s=2)7}=\beta_{2(s=1)7}+\Delta\beta_{2(s=2)7}=-0.163$ under $S_t=2$). 
The auto-regressive coefficients were larger under $S_t=2$ (as indicated with $\Delta\beta_{1(s=2)}$) except for the feeling that students used too much time for studying. This implied that persons who intended to quit had stronger auto-regressive coefficients and thus a more stable pattern (of the dysfunctional state variables like negative affective variables) over time (e.g. for \textit{afraid to fail}, $\beta_{1(s=1)7}=0.149$ under $S_t=1$ vs. $\beta_{1(s=2)3}=\beta_{1(s=1)3}+\Delta\beta_{1(s=2)3}=0.688$ under $S_t=2$). 
Under state $S_t=2$ (intention to quit) these interactions tended to be smaller in absolute size.

Figure~\ref{fig:pars3} includes the variances at the within and between level as well as the estimate for the switchback probability $P_{12}=P(S_{it} = 1| S_{i(t-1)} = 2)$. Within-level variances were similar and small for the first five scales (e.g., \textit{content not important} compared to the \textit{no PAP} and \textit{PAN} scales). This implied that time-specific variation was larger for these last two scales. The between-level variances were similar across the scales, showing that inter-individual differences were similar across all seven scales. The ICC for the within-level scales lay between 0.496 and 0.692. This implied that the scale levels had substantive inter-individual variation. Below, the forecast will take this aspect into account by using person-specific levels (random effects) for the individual forecast. The conditional switchback probability at each time point was estimated at $P_{12}=P(S_{it} = 1| S_{i(t-1)} = 2) = 0.097$. Accordingly, the conditional probability to maintain an intention to quit after having this intention at the first time is $P_{22}=P(S_{it} = 2| S_{i(t-1)} = 2) = 0.903$. Note that this probability is neither the overall probability for an intention to quit nor the same as the probability of the behavior to actually drop out (which was 36.1\%, see below).

Finally, Figure~\ref{fig:pars4} shows the estimates for the prediction of the latent discrete states (Markov Switching Model). This time-dependent switch was predicted primarily by \textit{PAN} (negative affect) and the scale \textit{being afraid to fail}. This indicates that the switch to an intention of quitting is associated particularly with negative affects and an expectation to fail the final exam. The remaining variables were less predictive for the discrete state change. In addition, the interactions between the cognitive skills (IQ) were close to zero.

\subsection{Results of the forecast: All measurement occasions}
In addition to the 36.1\% of persons who had actually quit the studies, a further 37.7\% showed a model-based state membership in the latent class $S_t=2$ (intention to quit) at the final time point $t=50$ (i.e., a total of 73.8\% of the students). The sensitivity of this intention for the actual observed drop-out was 0.86 and its specificity lay at 0.33. This of course, was expected because we assumed that more persons have an intention to quit than there are actually persons who drop-out (as a behavior). 

During the forecast period of 5 additional time points ($N_{t+}=5$) this percentage increased to 40.2\% (i.e., 3 more students were forecast to develop an intention to quit than at $t=50$). The average time point to switch from $S_t=1$ (no intention to quit) to $S_t=2$ (intention to quit) was at $t=24.6$ ($SD=8.9$). For those persons who later showed an actual drop-out, this switch occurred on average at $t=22.2$ ($SD=6.5$) which corresponds approximately to the 8th week of the math study program. This was considerably earlier than the actual drop-out was observed (on average at $t=45.0$, $SD=11.9$) which is approximately the 16th week of the math study program. Note that the period around $t=22.2$ is the critical period when the risk of dropping out of math studies becomes visible and potential interventions should be conducted then at the latest. This corresponds to a difference of 8 weeks before the actual behavior occurs. 

\begin{figure}
	\centering
	\includegraphics[width=\textwidth]{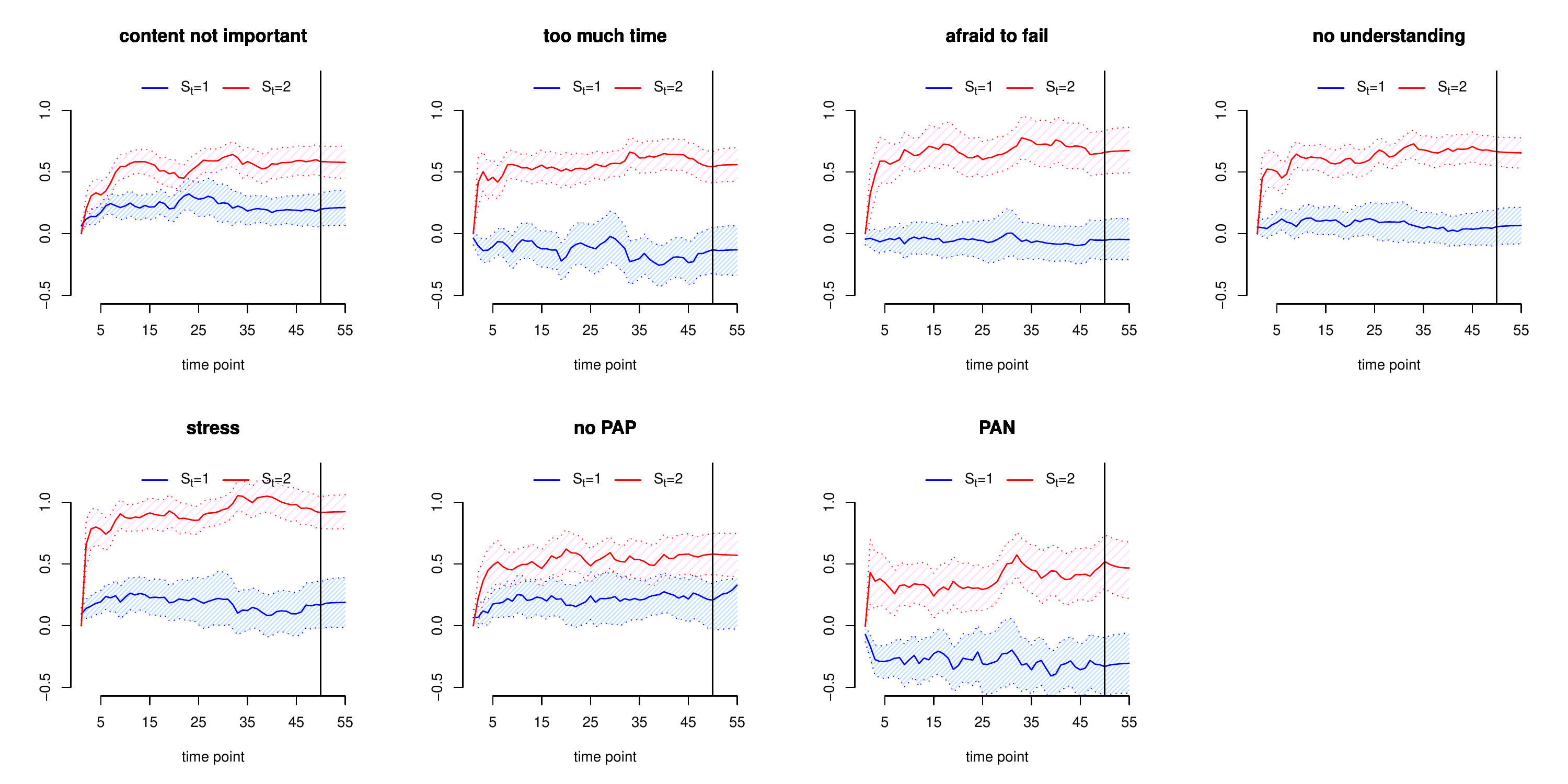}
	\caption{State-specific average trajectories and their forecast beyond $N_t=50$. $S_t=1$ (no intention to quit) and $S_t=2$ (intention to quit). Averages were taken at each time point based on the estimated membership in the discrete latent classes.}
	\label{fig:pred1}
\end{figure}

Figure~\ref{fig:pred1} shows the average trends of the seven within scales by the two discrete states $S_t=1$ (no intention to quit) and $S_t=2$ (intention to quit) separately. These scores were sampled from the posterior distribution using the estimated factor scores during time points 1 through $N_t=50$ and the forecast factor scores for the additional 5 time points ($N_{t+}=5$). The number of persons used to average these factor scores were different at each time point (estimates for class-membership were also drawn from the posterior distribution). Persons with an intention to quit ($S_t=2$) exhibited higher average scores particularly in the scales ``too much time'', ``afraid to fail'', ``no understanding'', ``stress'' and ``PAN''. The scores for ``no PAP'' were overlapping and the scores for the scale ``content not important'' were slightly higher for persons with an intention to quit.

\begin{figure}
	\centering
	\includegraphics[width=.75\textwidth]{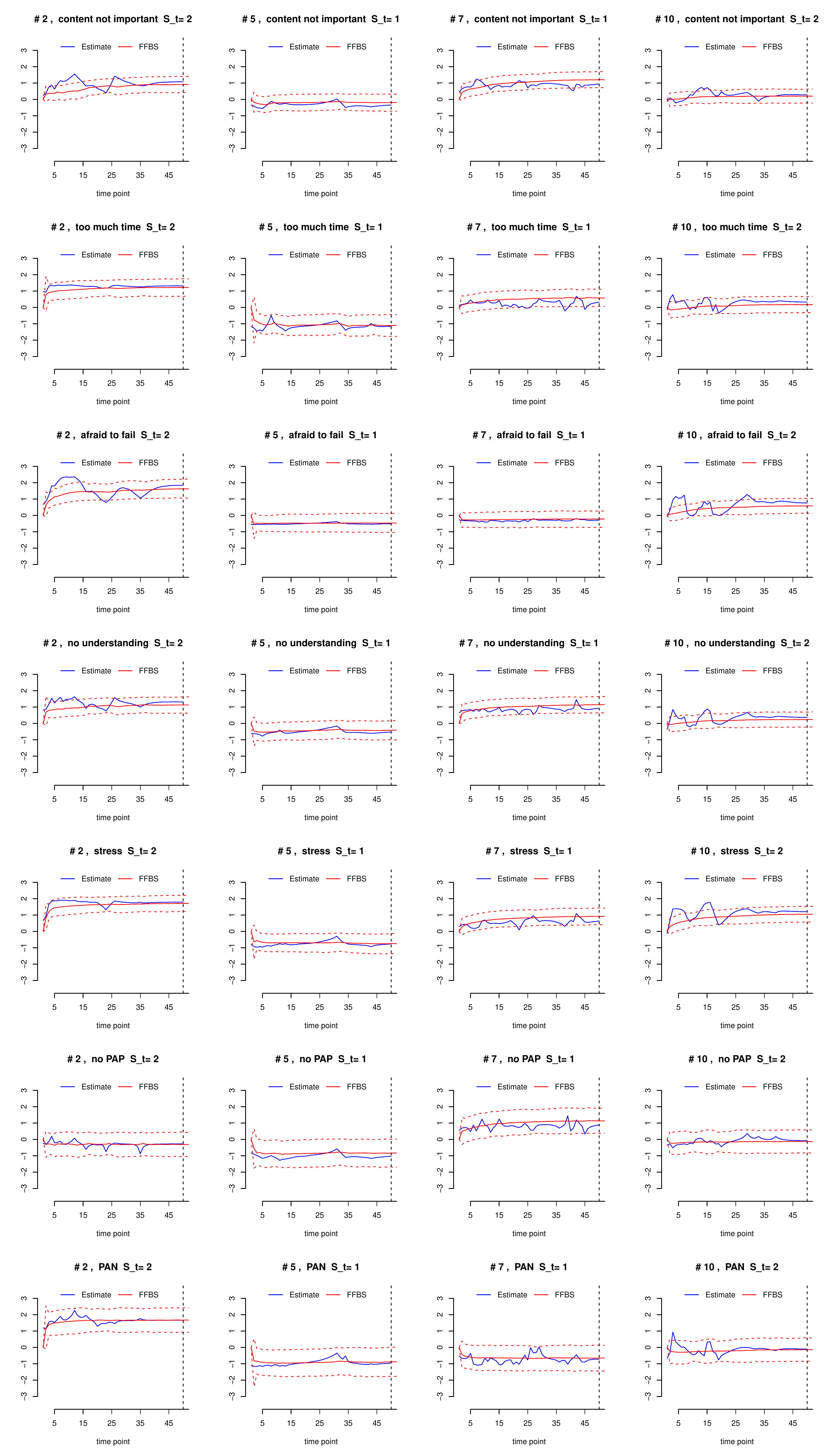}
	\caption{Individual trajectories for four random students (\#2, \#5, \#7, \#10; as columns). Blue indicates the factor score estimate based on all measurement occasions. Red indicates the forecast ($t>50$) and the smoothing ($t<50$). $S_t$ indicates the final membership in the discrete state (intention to quit) at $t=55$.}
	\label{fig:ffbs1}
\end{figure}

Figure~\ref{fig:ffbs1} illustrates individual trajectories of four randomly selected students (Each column represents one student. The rows are the dependent variables). The smooth red lines show how the FFBS algorithm recovers the actual trajectories based on the factor scores (blue). For the additional time points beyond $N_t=50$, the forecast illustrates the future development of the students. Note how students \#2 (first column) and \#10 (last column) in particular show different trajectories than students \#5 (second column) and \#7 (third column) in the scale ``afraid to fail'' and ``PAN'' which are indicative for their switch to the discrete state $S_t=2$.

\subsection{Results of the forecast: Half of the measurement occasions}
\begin{figure}
	\centering
	\includegraphics[angle=00,width=.5\textwidth]{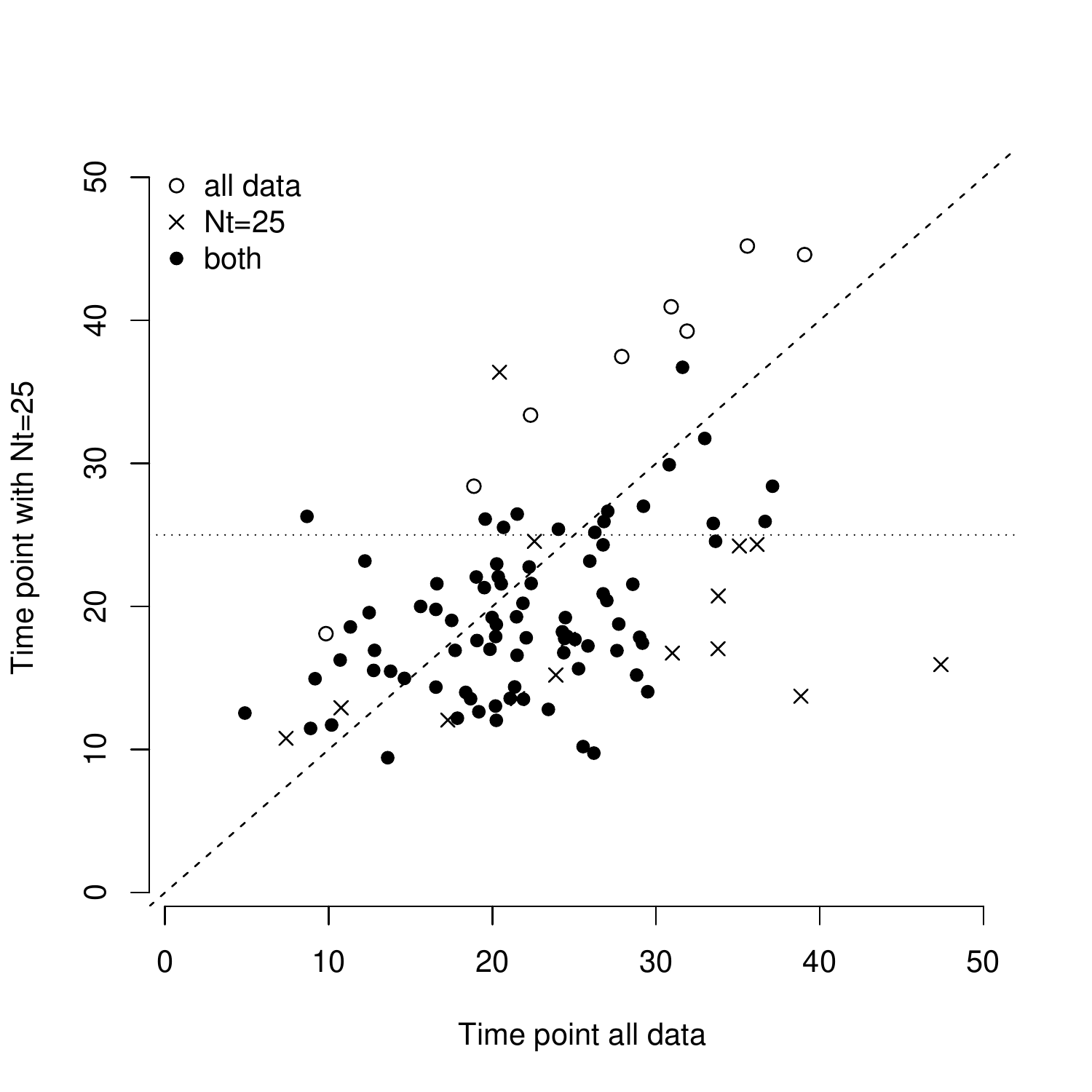}
	\caption{Estimated time points for the switch from the discrete latent state $S_t=1$ to $S_t=2$ based on the complete sample ("all data") in comparison to the reduced data set with $N_t=25$ (and a forecast of $N_{t+}=25$). This dashed line is a diagonal; the dotted line indicates where the forecast for the reduced data set starts.}
	\label{fig:switchest}
\end{figure}

In a second experiment, we used data only from the first $N_t=25$ measurement occasions. We then forecast an additional $N_{t+}=25$ measurement occasions. With this experiment, we wanted to test individual factor scores that were forecast in comparison to the actual factor scores estimated based on all measurement occasions. The forecast of the membership of the discrete latent class at $t=50$ was very similar to the one obtained from the whole sample used above with an overlap of 81\% identical classifications in $S_{50}$. The average time point of switching was $23.1$ ($SD=10.1$); the correlation between the time points from this reduced sample and those estimated from the full sample was 0.62. Figure~\ref{fig:switchest} illustrates this relationship.

\begin{figure}
	\centering
	\includegraphics[angle=0,width=.75\textwidth]{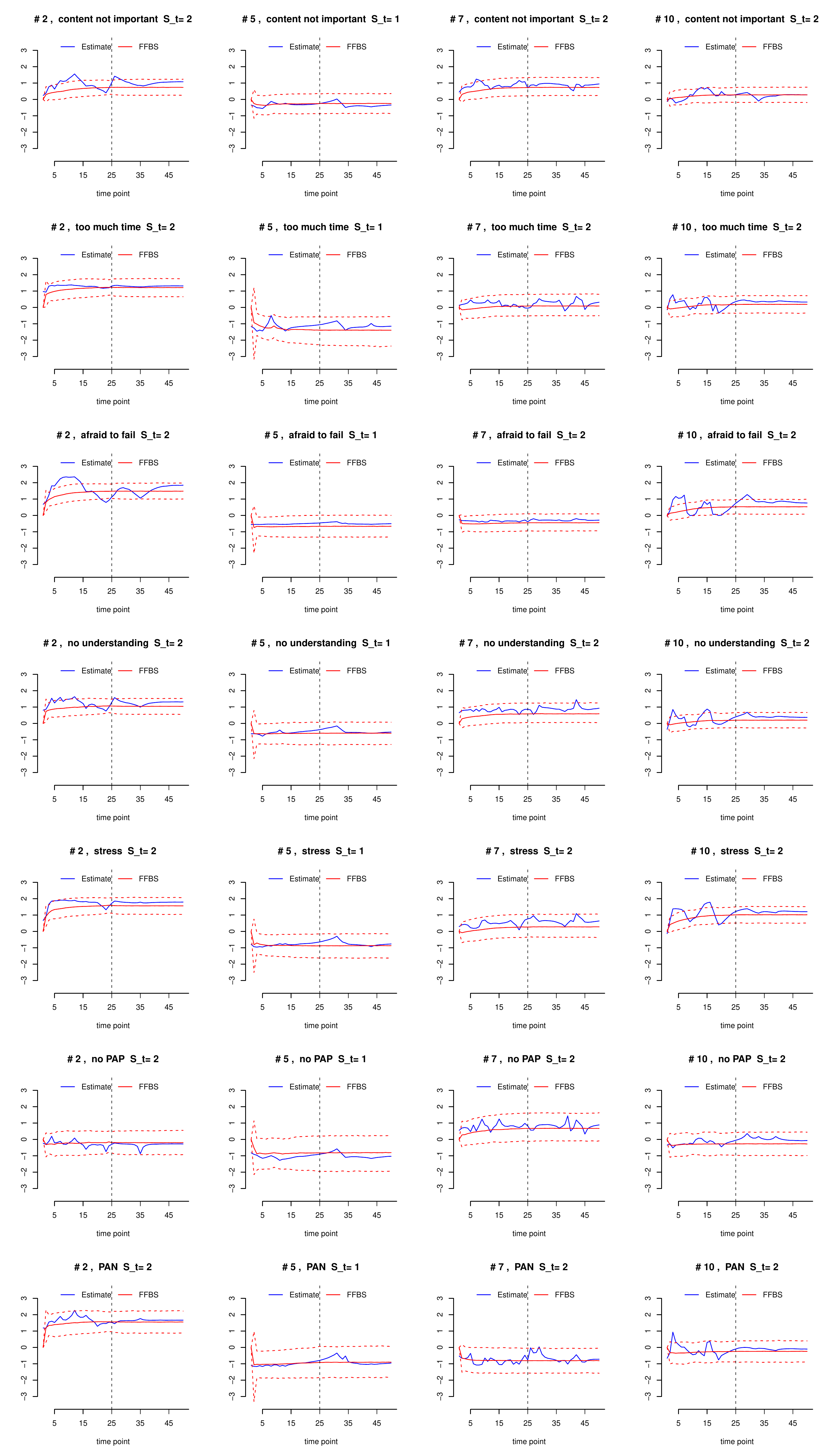}
	\caption{Individual trajectories for four random students (\#2, \#5, \#7, \#10; as columns). Blue indicates the factor score estimate based on all measurement occasions. Red indicates the forecast with data only for the initial 25 measurement occasions (dashed vertical line). $S_t$ indicates the final membership in the discrete state (intention to quit) at $t=50$.}
	\label{fig:ffbs2}
\end{figure}

Figure~\ref{fig:ffbs2} shows this forecast for all seven scales for a random selection of four students (same as above). The factor score estimates obtained from the original complete sample (with all measurement occasions) are shown in blue. They are used as a reference to judge the forecast (red). The forecast intervals (FI) include these factor scores reliably. As is typical for the FFBS method, the method is used as forecast for data points $N_{t+}$, and it smoothes over the actual data for the initial time points $1,\ldots N_t$ as can be seen nicely in Figure~\ref{fig:ffbs2} (see also West \& Harrison, 1997; Petris et al., 2009). Thus, the estimates between $t=1,\ldots 25$ are smoothed predictions of the actual factor scores. Note that the prediction takes the inter-individual different levels directly into account using the random effect estimates. 

\section{Simulation Study}
In this section, we present a small simulation study that evaluates the performance of the model specified within the NDLC-SEM framework for the purpose of forecasting based on the FFBS method as described above. We restrict our simulation study to the model specified in the example section (abbreviated NDLC-SEM). We investigate how the forecast depends on sample size ($N_1$) and number of measurement occasions ($N_t$).

\subsection{Data generation}
Data were generated according to the model presented in Equations~\eqref{eqex:l1:m1} to \eqref{eqex:markov:NDLC-SEM3}. In contrast to the empirical example, we used only three latent factors instead of seven (due to computational constraints). For each factor, three observed variables were generated. Population level parameters are based on the estimates obtained from the empirical example. We calculated means and standard deviations for each parameter across the seven scales we had analyzed in the empirical data (see Table~\ref{tab:simdata} and Figure~\ref{fig:simdata} in the Appendix). Parameters for each replication were randomly sampled from normal distributions using these population values. This allowed us to consider more general model specifications that include the different scales from the example but might be generalizable to other scales (that follow similar dynamic patterns).

Data were generated for $N_1=25$ vs. $N_1=50$ (individuals; e.g. students) as well as $N_t=25$ vs. $N_t=50$ measurement occasions. Data for additional $N_{t+}=10$ measurement occasions were generated and used to evaluate model performance after model estimation (see details below).

\subsection{Model estimation and evaluation}
The NDLC-SEM and its forecast were implemented according to the description provided above with priors given in Equations~\eqref{eq:prioralpha} to \eqref{eq:prioromega}. A forecast for the additional $N_{t+}=10$ time points after $N_t$ was conducted using the FFBS method. The NDLC-SEM was run with 2 chains and 10,000 iterations each with 5,000 iterations burn-in. A total of $R=100$ data sets were generated under each condition. Convergence was checked for each replication using the Rhat statistic (see above).

We used four major outcomes to evaluate the performance of the models. First, we investigated the \textit{sensitivity and specificity} of the NDLC-SEM for the state extraction with regard to the actual discrete state membership generated for the data. We distinguished between overall measures, for the state estimation during the first $N_t$ measurement occasions, and the state forecast for the $h=1\ldots N_{t+}$ time points Second, we calculated the average \textit{$95\%$ coverage rates} for the factor scores based on the forecasting intervals for the forecast factor score estimates. Third, we used a \textit{score function $\delta_{h}$} to evaluate how well the forecast of the latent within factors replicated the actual factor scores of the three latent within factors by
\begin{equation}
	\delta_h=\sum_i\sum_j \left(\hat{\eta}_{ihj}-\eta_{ihj}\right)^2, \quad i=1\ldots N_1, j=1\ldots 3
\end{equation}
for each forecast time point $h=1\ldots N_{t+}$ (cf. Gneiting, 2011). Finally, we investigated how precise the forecast was by calculating the \textit{average width of the forecasting intervals} (FI) for the forecast factor score estimates at each forecast time point.

\subsection{Results}

\paragraph{State prediction}
Table~\ref{tab:simsens} shows the results for the state estimation and the state prediction ($S_{it}$). Sensitivity for the discrete state extraction was above 0.91 under all conditions. This indicated that persons who actually switched could reliably be detected. This also extended to the forecast of the discrete states. Specificity was lower with values between 0.83 and 0.88 for the time points in the interval 1 through $N_t$, that is, the prediction of the state membership for the observed data. Specificity was lower for the forecast time points with values ranging between 0.53 and 0.70; they were lower for smaller sample size $N_1=25$ compared to $N_1=50$. This was due to the fact that too many persons were forecast to switch to the latent state $S=2$ in this interval. It implied that in order to achieve acceptable specificity rates, at least $N_1=50$ individuals should be included.

\begin{table}
	\centering
	\caption{Sensitivity and specificity for the latent class extraction using the NDLC-SEM and 95\% coverage rates for the forecast intervals. Overall includes all time points, Observed data is restricted to the time points 1 through $N_t$, and Forecast is restricted to the forecast time points $h=1\ldots N_{t+}$.}
	\begin{tabular}{cc ccc ccc c}
		\hline
		&&\multicolumn{3}{c}{Sensitivity} & \multicolumn{3}{c}{Specificity} & Coverage\\
		$N_t$ &$N_1$& Overall & Observed data & Forecast & Overall & Observed data & Forecast  \\ 
		\hline
		25 & 25 & 0.92 & 0.91 & 0.93 & 0.83 & 0.85 & 0.62 & 0.90\\ 
		& 50 & 0.93 & 0.93 & 0.94 & 0.87 & 0.88 & 0.70 & 0.88\\ 
		50 & 25 & 0.96 & 0.96 & 0.95 & 0.81 & 0.83 & 0.53 & 0.89\\ 
		& 50 & 0.94 & 0.95 & 0.94 & 0.85 & 0.86 & 0.70 & 0.88\\ 
		\hline\hline
	\end{tabular}
	\label{tab:simsens}
\end{table}

\paragraph{Coverage rates}
The last column in Table~\ref{tab:simsens} shows the 95\% coverage rates for the forecast intervals averaged across time points, persons, and the three latent factors for the NDLC-SEM. Across all conditions, coverage rates were between 88\% and 90\%. The NDLC-SEM thus showed somewhat lower coverage rates than the nominal 95\% level. This finding can be attributed to the lower specificity described above, that is, persons were classified too often to switch to the discrete state $S=2$ (i.e., intention to quit). 

\begin{figure}
	\centering
	\includegraphics[width=\textwidth]{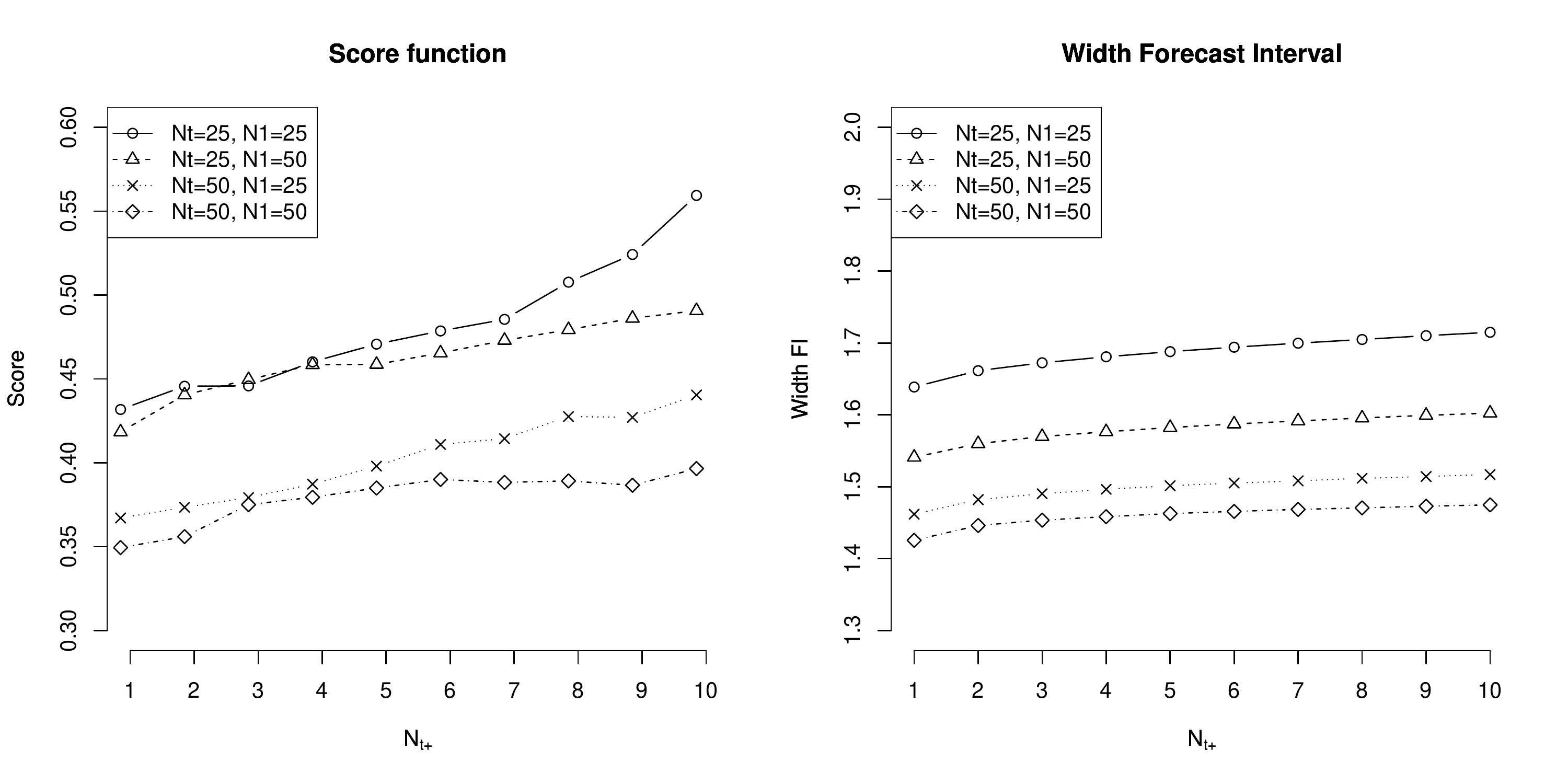}
	\caption{Left: (Quadratic) score function under the different conditions of sample size and time points across the forecast time points. Right: Average width of the forecast intervals (FI) under the different conditions of sample size and time points across the forecast time points}
	\label{fig:simscore}
\end{figure}

\paragraph{Score function}
The left hand panel in Figure~\ref{fig:simscore} illustrates the (quadratic) score function that indicates the average quadratic difference between the actual factor scores and the forecast factor scores. It provides information about how precise the forecast was under the different conditions and how it changed with more forecast time points. The score function was mostly affected by the number of measurement occasion ($N_t$), resulting in similar forecast precision for both sample sizes under the condition of $N_t=25$ vs. $N_t=50$. Under $N_t=50$, the score function resulted in considerably smaller values (for both sample sizes $N_1=25$ and $N_1=50)$.

\paragraph{Width of the forecast interval (FI)}
The right panel in Figure~\ref{fig:simscore} shows the average width of the FI across the forecast time points. Across all conditions, the width of the FI increased with additional time points that were forecast (looking like a megaphone). As expected, the FI width decreased with more time points $N_t$ and larger sample sizes $N_1$.

\section{Discussion}
The collection and availability of ILD has changed the view of relationships between different psychological constructs as well as dynamic changes taking place within them. The examination of ILD offers detailed insights into the temporal changes of psychological phenomena and the way they fluctuate. In addition, from a theoretical perspective, conceptually distinct levels of psychological phenomena can be separated and analyzed. In our university drop out study example, we were interested in the separation of stable a priori inter-individual differences (e.g., cognitive abilities, school achievement, personality factors), intra-individual changes (e.g., affective states, dysfunctional cognitions), and unobserved heterogeneity (e.g., pre-decisional discrete states of intentions to quit). We were particularly interested in the identification of individuals showing critical states on intra-individual variables or discrete latent states that indicate potential drop out.

In order to examine this substantive problem, we used dynamic latent variable modeling (e.g., Asparouhov, et al., 2017, 2018) and Bayesian estimation (Gelman, Carlin, Stern, Dunson, Vehtari, \& Rubin, 2013), specifically the NDLC-SEM framework (Kelava \& Brandt, 2019) which is capable of addressing these different conceptual aspects within one overall model. Within this approach, we specified a model that is capable of dealing with these different data-levels. In a dynamic autoregressive model, we describe intra-individual changes of seven within-level latent variables (e.g., subjective importance of content, stress, affective states etc.). Furthermore, we examined the interaction between these within-level latent state variables and the between-level students’ cognitive abilities variable, in order to explain the transitions to discrete latent states that represent intentions to quit math studies.

Building on previous work of West and Harrison (1997), we presented an adaptation of a Forward Filtering Backward Sampling (FFBS) method that we used for our forecasting problem of time-dependent latent states. We described so-called observation and evolution equations, priors, state, forecast distributions, updating, and posterior probabilities which allow the prediction of states as well as the quantification of uncertainty. The FFBS procedure gives forecasts of both time-dependent continuous latent factors (e.g., affective states) and time-dependent discrete latent states (e.g., intention to quit).

In the empirical example, we were able to show that the development of an intention to drop out is related to baseline and time-specific variables in a complex fashion. We used this information to identify those persons who intend to drop out. We were able to show that the negative affects expressed in particular are predictive for persons to develop such an intention. In addition, the individuals in these two states (intention/no intention) show a very distinct response pattern on the affective states -- they are afraid to fail, show no understanding of the content, believe that they invest to much time in studying, feel stressed and, again, show negative affective states. These variables can be helpful in determining candidates for drop-out.

We further showed that the FFBS is a valid method for forecasting individual trajectories and changes over time. Some initial measures, or a part of a time-series (e.g., the first half of a semester) can be used as a basis to reliably forecast how students will develop with regard to the affective scales. This information is useful to make predictions early on about who might develop an intention to drop out. 
In addition, we were able to show that the forecast is effective even if only the first half of the measurement occasions is available. 

The forecasting demonstrated how the substantial problem of student drop out from university math may be monitored from an early stage, and the findings point to time slots for successful interventions. Given that negative affect and fear of failure were shown to increase the likelihood of intending to quit, such interventions might usefully be designed to address negative emotions, for example by offering emotional support to students. Since the affective changes must also be an expression of an overload of professional/cognitive skills, it obviously remains reasonable to provide extra tutorials to help with following the course content.
Our findings are also helpful in indicating when such an intervention might be most helpful: the changes in emotional states related to intention to quit occurred, on average, around eight weeks before any actual dropout, suggesting that early monitoring and early interventions seem more likely to be successful. However, we would like to note that, in this paper, we focused on forecasting in the context of ILD, but we did not actually implement an intervention. Instead, we have demonstrated how the forecasting of critical states in such a complex arrangement of data (levels) could form the basis of a time-sensitive intervention.

The results of the simulations study show that the sensitivity of the proposed model is very good, even for small sample sizes ($N_1 =25$) and a low number of measurements ($N_t =25$) when forecasting of additional $N_{t+}=10$ time points is conducted. Specificity is acceptable for low sample sizes $N_1=25$, indicating a progressive classification of students at risk (which is acceptable from a substantive point of view). However, specificity improves substantially for sample sizes of $N_1=50$, reaching satisfactory levels. Note that in our empirical example, we had a larger sample size of $N_1=122$. In order to quantify the precision of our forecasts, we examined the average quadratic difference between the forecast and the true values of the latent variables using a score function. Results of both the score function and forecasting intervals showed that the NDLC-SEM model was more precise when the number of prior time points (before forecasting) was large ($N_t=50$). As analytically expected, the forecasts are less precise with increasing distance/number of predicted time points (i.e., megaphone shape of forecasting interval). The coverage rates were approx. 90\% across all conditions.


\subsection{Limitations}
There are several limitations that need to be reflected on in the context of this paper. First, generally speaking, little is known about the finite sample properties of dynamic latent variable frameworks with respect to the stability of estimates on the different data levels. Systematic simulation studies that examine the balance of the number of individuals (N) vs. number of measurement occasions in dynamic models are rare (e.g., Schultzberg, M., \& Muthén, B., 2018). This has implications for the quality of forecasting in this specific context.

Second, we have focused on the Bayesian perspective of forecasting. We did not apply frequentist ways of forecasting. For example, the Kalman filter (Kalman, 1960) is a standard technique of forecasting in state space modeling and time series modeling in general. The reason is the widespread availability of Bayesian MCMC sampling in software packages. In addition, an implementation of a computationally efficient frequentist estimation method for dynamic latent variable models is not straightforward (due to the high-dimensional numeric optimization requirements). This holds especially for these complex time series with several data levels. Bayesian methods are particularly helpful when using information from previous time points or cohorts/samples.

Third, model fit and stability of the model predictions is a critical issue. At this point of research, no global model fit or model fit indices (such as CFI versions for the Bayesian estimator) are available. New research proposed Bayesian adaptations of these fit indices, but they have only been used for very simple CFA models (e.g., Garnier-Villarreal, \& Jorgensen, in press). Information criteria that can be used to compare models have also not been systematically investigated for latent variable models with latent dynamic (discrete) states.

Fourth, there might be other approaches to time series that have their origin in machine learning. For example, further developments of Restricted Boltzmann Machines (RBM), like the conditional RBM (cRBM) (Tayler, Hinton, \& Roweis, 2006) or, similarly, the temporal RBM (Sutskever, \& Hinton, 2007) include temporal dependencies. Convolutional layers show promising results in modeling time-series by including time-lagged information by 1-dimensional convolutions (e.g., convRBM; Lee, Grosse, Ranganath, \& Ng, 2009), as well as recurrent neural networks (RNNs) with long-short term memory cells (LSTM; Hochreiter \& Schmidhuber, 1997). However, to our knowledge, these methods either do not consider different data levels, or they are not typically applied to small data sets (e.g., L\"angkvist, Karlsson, \& Loutfi, 2014). Thus, open research questions remain concerning the suitability of these methods for the sample sizes that can typically be obtained in psychological research. The potential of these methods in this context is unknown and needs examination.

Finally, test motivation and the willingness to respond to items which are presented repeatedly as part of an ILD assessment are important issues. In our SAM study, we decided to use a small number of items and subsets of scales that were presented three times a week. This had the negative effect of lowering the reliability of the scales which in turn leads to increased variability of the parameter estimates of the effects of constructs of interest. However, we believe that the quality of the data was increased substantially as a result of this decision.


\subsection{Future Directions and open questions}
A number of questions remain open and could be addressed in future research.

First, it is unclear how well dynamic SEM will perform compared with techniques from statistical learning that allow for classification and prediction of critical states (see above). Important aspects will be sample size requirements and the complexity of the models (e.g., for both dynamic SEM and machine learning techniques).

Second, since model complexity is important, regularization techniques (both Bayesian and frequentist, e.g. Hastie et al., 2009) will be another aspect that will influence forecasting in models suitable for ILD. Depending on the possibilities to reduce the complexities, alternative procedures might arise which have not yet been extensively discussed in psychometrics. Regularization is important because ILD have substantially lower sample sizes than data that are used in standard machine learning situations.

Third, a combination of techniques developed in the field of psychometrics with techniques from the field of machine learning seems to be very promising. It will be important to combine their strengths (e.g., the sparsity of psychometric models and their causal orientation with the precision of neural networks).  We believe that a new important field of research will emerge with new perspectives on forecasting in ILD.

\vspace{\fill}\clearpage


\vspace{\fill}

\newpage
\appendix

\section{Additional results from the empirical example}

\begin{table}[ht]
	\centering
	\caption{Parameter estimates for the latent discrete State $S_t=1$ (Mean, SD, 2.5\% and 97.5\% Percentiles of the Posterior distribution and the Rhat statistic).}
	\begin{tabular}{rrrrrr}
		\hline
		& Mean & SD & 2.5\% & 97.5\% & Rhat \\ 
		\hline
		$\alpha_{21(s=1)1}$  & 0.09 & 0.03 & 0.02 & 0.16 & 1.00 \\ 
		$\alpha_{21(s=1)2}$  & -0.06 & 0.05 & -0.16 & 0.03 & 1.00 \\ 
		$\alpha_{21(s=1)3}$  & -0.06 & 0.04 & -0.14 & 0.03 & 1.01 \\ 
		$\alpha_{21(s=1)4}$  & 0.07 & 0.05 & -0.02 & 0.16 & 1.02 \\ 
		$\alpha_{21(s=1)5}$  & 0.13 & 0.04 & 0.04 & 0.22 & 1.00 \\ 
		$\alpha_{21(s=1)6}$  & 0.12 & 0.04 & 0.03 & 0.21 & 1.01 \\ 
		$\alpha_{21(s=1)7}$  & -0.18 & 0.06 & -0.29 & -0.06 & 1.02 \\ 
		$\beta_{2(s=1)1}$    & -0.18 & 0.09 & -0.35 & -0.01 & 1.00 \\ 
		$\beta_{2(s=1)2}$    & -0.87 & 0.15 & -1.18 & -0.60 & 1.00 \\ 
		$\beta_{2(s=1)3}$    & -0.77 & 0.12 & -1.02 & -0.56 & 1.03 \\ 
		$\beta_{2(s=1)4}$    & -0.75 & 0.14 & -1.02 & -0.49 & 1.01 \\ 
		$\beta_{2(s=1)5}$    & -0.81 & 0.13 & -1.09 & -0.57 & 1.01 \\ 
		$\beta_{2(s=1)6}$    & -0.22 & 0.12 & -0.46 & 0.00 & 1.01 \\ 
		$\beta_{2(s=1)7}$    & -1.12 & 0.17 & -1.45 & -0.79 & 1.01 \\ 
		$\beta_{1(s=1)1}$    & 0.49 & 0.04 & 0.40 & 0.57 & 1.07 \\ 
		$\beta_{1(s=1)2}$    & 0.33 & 0.04 & 0.25 & 0.41 & 1.02 \\ 
		$\beta_{1(s=1)3}$    & 0.15 & 0.04 & 0.06 & 0.24 & 1.07 \\ 
		$\beta_{1(s=1)4}$    & 0.22 & 0.05 & 0.12 & 0.32 & 1.02 \\ 
		$\beta_{1(s=1)5}$    & 0.38 & 0.04 & 0.30 & 0.45 & 1.01 \\ 
		$\beta_{1(s=1)6}$    & 0.32 & 0.05 & 0.22 & 0.41 & 1.01 \\ 
		$\beta_{1(s=1)7}$    & 0.28 & 0.04 & 0.20 & 0.35 & 1.01 \\ 
		$\omega_{2(s=1)1}$   & 0.34 & 0.11 & 0.15 & 0.58 & 1.12 \\ 
		$\omega_{2(s=1)2}$   & 0.11 & 0.09 & -0.07 & 0.29 & 1.01 \\ 
		$\omega_{2(s=1)3}$   & -0.30 & 0.10 & -0.50 & -0.11 & 1.04 \\ 
		$\omega_{2(s=1)4}$   & 0.17 & 0.10 & -0.03 & 0.37 & 1.01 \\ 
		$\omega_{2(s=1)5}$   & -0.11 & 0.07 & -0.27 & 0.03 & 1.00 \\ 
		$\omega_{2(s=1)6}$   & 0.46 & 0.13 & 0.21 & 0.72 & 1.01 \\ 
		$\omega_{2(s=1)7}$   & -0.18 & 0.08 & -0.34 & -0.02 & 1.05 \\ 
		\hline
	\end{tabular}
\end{table}

\begin{table}
	\centering
	\caption{Parameter estimates for the latent discrete State $S_t=2$ (Mean, SD, 2.5\% and 97.5\% Percentiles of the Posterior distribution and the Rhat statistic).}
	\begin{tabular}{rrrrrr}
		\hline
		& Mean & SD & 2.5\% & 97.5\% & Rhat \\ 
		\hline
		$\Delta\alpha_{21(s=2)1}$ & 0.14 & 0.03 & 0.09 & 0.20 & 1.03 \\ 
		$\Delta\alpha_{21(s=2)2}$ & 0.51 & 0.04 & 0.44 & 0.59 & 1.03 \\ 
		$\Delta\alpha_{21(s=2)3}$ & 0.28 & 0.03 & 0.23 & 0.34 & 1.01 \\ 
		$\Delta\alpha_{21(s=2)4}$ & 0.38 & 0.04 & 0.30 & 0.46 & 1.03 \\ 
		$\Delta\alpha_{21(s=2)5}$ & 0.38 & 0.04 & 0.30 & 0.45 & 1.01 \\ 
		$\Delta\alpha_{21(s=2)6}$ & 0.18 & 0.04 & 0.11 & 0.26 & 1.03 \\ 
		$\Delta\alpha_{21(s=2)7}$ & 0.32 & 0.05 & 0.23 & 0.41 & 1.03 \\ 
		$\Delta\beta_{2(s=2)1}$   & -0.05 & 0.08 & -0.22 & 0.11 & 1.05 \\ 
		$\Delta\beta_{2(s=2)2}$   & -0.05 & 0.13 & -0.30 & 0.23 & 1.03 \\ 
		$\Delta\beta_{2(s=2)3}$   & 0.32 & 0.11 & 0.13 & 0.54 & 1.05 \\ 
		$\Delta\beta_{2(s=2)4}$   & 0.05 & 0.13 & -0.20 & 0.31 & 1.07 \\ 
		$\Delta\beta_{2(s=2)5}$   & 0.19 & 0.12 & -0.06 & 0.42 & 1.04 \\ 
		$\Delta\beta_{2(s=2)6}$   & 0.01 & 0.12 & -0.22 & 0.24 & 1.03 \\ 
		$\Delta\beta_{2(s=2)7}$   & 0.95 & 0.15 & 0.68 & 1.25 & 1.04 \\ 
		$\Delta\beta_{1(s=2)1}$   & 0.16 & 0.04 & 0.08 & 0.24 & 1.02 \\ 
		$\Delta\beta_{1(s=2)2}$   & -0.02 & 0.05 & -0.12 & 0.09 & 1.02 \\ 
		$\Delta\beta_{1(s=2)3}$   & 0.54 & 0.05 & 0.45 & 0.62 & 1.11 \\ 
		$\Delta\beta_{1(s=2)4}$   & 0.16 & 0.05 & 0.06 & 0.27 & 1.03 \\ 
		$\Delta\beta_{1(s=2)5}$   & 0.10 & 0.04 & 0.01 & 0.18 & 1.02 \\ 
		$\Delta\beta_{1(s=2)6}$   & 0.20 & 0.05 & 0.10 & 0.30 & 1.02 \\ 
		$\Delta\beta_{1(s=2)7}$   & 0.35 & 0.04 & 0.26 & 0.43 & 1.03 \\ 
		$\Delta\omega_{2(s=2)1}$  & -0.27 & 0.10 & -0.47 & -0.09 & 1.03 \\ 
		$\Delta\omega_{2(s=2)2}$  & 0.46 & 0.13 & 0.22 & 0.73 & 1.02 \\ 
		$\Delta\omega_{2(s=2)3}$  & 0.31 & 0.08 & 0.16 & 0.48 & 1.01 \\ 
		$\Delta\omega_{2(s=2)4}$  & 0.08 & 0.11 & -0.12 & 0.30 & 1.04 \\ 
		$\Delta\omega_{2(s=2)5}$  & 0.31 & 0.10 & 0.13 & 0.51 & 1.00 \\ 
		$\Delta\omega_{2(s=2)6}$  & -0.41 & 0.14 & -0.69 & -0.12 & 1.02 \\ 
		$\Delta\omega_{2(s=2)7}$  & 0.45 & 0.13 & 0.21 & 0.73 & 1.04 \\ 
		\hline
	\end{tabular}
\end{table}

\begin{table}
	\centering
	\begin{tabular}{rrrrrr}
		\hline
		& Mean & SD & 2.5\% & 97.5\% & Rhat \\ 
		\hline
		$\sigma^2_{\zeta_{11}}$ & 0.06 & 0.00 & 0.05 & 0.06 & 1.00 \\ 
		$\sigma^2_{\zeta_{12}}$ & 0.08 & 0.01 & 0.07 & 0.09 & 1.01 \\ 
		$\sigma^2_{\zeta_{13}}$ & 0.05 & 0.00 & 0.05 & 0.06 & 1.00 \\ 
		$\sigma^2_{\zeta_{14}}$ & 0.07 & 0.01 & 0.06 & 0.08 & 1.00 \\ 
		$\sigma^2_{\zeta_{15}}$ & 0.07 & 0.00 & 0.06 & 0.08 & 1.01 \\ 
		$\sigma^2_{\zeta_{16}}$ & 0.16 & 0.01 & 0.14 & 0.19 & 1.00 \\ 
		$\sigma^2_{\zeta_{17}}$ & 0.17 & 0.01 & 0.15 & 0.19 & 1.02 \\ 
		$\sigma^2_{\zeta_{21}}$ & 0.09 & 0.01 & 0.07 & 0.12 & 1.00 \\ 
		$\sigma^2_{\zeta_{22}}$ & 0.16 & 0.02 & 0.12 & 0.21 & 1.00 \\ 
		$\sigma^2_{\zeta_{23}}$ & 0.12 & 0.02 & 0.09 & 0.16 & 1.00 \\ 
		$\sigma^2_{\zeta_{24}}$ & 0.13 & 0.02 & 0.10 & 0.17 & 1.00 \\ 
		$\sigma^2_{\zeta_{25}}$ & 0.13 & 0.02 & 0.10 & 0.17 & 1.00 \\ 
		$\sigma^2_{\zeta_{26}}$ & 0.16 & 0.02 & 0.12 & 0.21 & 1.00 \\ 
		$\sigma^2_{\zeta_{27}}$ & 0.19 & 0.03 & 0.14 & 0.25 & 1.00 \\ 
		$\sigma^2_{\zeta_{3}}$ & 0.20 & 0.04 & 0.14 & 0.28 & 1.01 \\ 
		$P_{12}$ & 0.10 & 0.00 & 0.09 & 0.10 & 1.00 \\ 
		\hline
	\end{tabular}
\end{table}

\begin{table}
	\centering
	\caption{Parameter estimates for the Markov Switching Model (Mean, SD, 2.5\% and 97.5\% Percentiles of the Posterior distribution and the Rhat statistic).}
	\begin{tabular}{lrrrrr}
		\hline
		& Mean & SD & 2.5\% & 97.5\% & Rhat \\ 
		\hline
		$\gamma_{2}$ & -1.17 & 0.79 & -2.67 & 0.42 & 1.03 \\ 
		$\gamma_{31}$& -0.10 & 0.34 & -0.77 & 0.58 & 1.01 \\ 
		$\gamma_{32}$& 0.26 & 0.33 & -0.40 & 0.91 & 1.02 \\ 
		$\gamma_{33}$& -0.83 & 0.48 & -1.77 & 0.12 & 1.03 \\ 
		$\gamma_{34}$& 0.48 & 0.55 & -0.58 & 1.54 & 1.03 \\ 
		$\gamma_{35}$& -0.56 & 0.42 & -1.41 & 0.23 & 1.03 \\ 
		$\gamma_{36}$& 0.11 & 0.25 & -0.38 & 0.58 & 1.01 \\ 
		$\gamma_{37}$& -0.71 & 0.21 & -1.13 & -0.31 & 1.03 \\ 
		$\gamma_{41}$ & -0.74 & 0.68 & -2.10 & 0.54 & 1.06 \\ 
		$\gamma_{42}$ & -0.19 & 0.76 & -1.73 & 1.24 & 1.02 \\ 
		$\gamma_{43}$ & -0.86 & 0.74 & -2.32 & 0.57 & 1.04 \\ 
		\hline
	\end{tabular}
\end{table}

\begin{table}
	\centering
	\caption{Parameter estimates for the Factor Loadings (Mean, SD, 2.5\% and 97.5\% Percentiles of the Posterior distribution and the Rhat statistic).}
	\begin{tabular}{rrrrrr}
		\hline
		& Mean & SD & 2.5\% & 97.5\% & Rhat \\ 
		\hline
		$\lambda_{101}$ & 1.31 & 0.03 & 1.25 & 1.36 & 1.00 \\ 
		$\lambda_{102}$ & 0.92 & 0.03 & 0.86 & 0.97 & 1.00 \\ 
		$\lambda_{103}$ & 0.92 & 0.03 & 0.87 & 0.97 & 1.01 \\ 
		$\lambda_{104}$ & 0.89 & 0.03 & 0.83 & 0.95 & 1.01 \\ 
		$\lambda_{105}$ & 1.10 & 0.02 & 1.06 & 1.14 & 1.00 \\ 
		$\lambda_{106}$ & 1.01 & 0.02 & 0.97 & 1.06 & 1.00 \\ 
		$\lambda_{201}$ & 1.20 & 0.23 & 0.76 & 1.67 & 1.01 \\ 
		$\lambda_{202}$ & 0.41 & 0.22 & 0.04 & 0.87 & 1.01 \\ 
		\hline
	\end{tabular}
\end{table}

\begin{table}
	\centering
	\caption{Parameter estimates for the Residual variances of the Items (Mean, SD, 2.5\% and 97.5\% Percentiles of the Posterior distribution and the Rhat statistic).}
	\begin{tabular}{rrrrrr}
		\hline
		& Mean & SD & 2.5\% & 97.5\% & Rhat \\ 
		\hline
		$\sigma_{\epsilon^2_{11}}$ & 0.35 & 0.01 & 0.32 & 0.37 & 1.00 \\ 
		$\sigma_{\epsilon^2_{12}}$ & 0.22 & 0.01 & 0.20 & 0.24 & 1.00 \\ 
		$\sigma_{\epsilon^2_{13}}$ & 0.58 & 0.02 & 0.55 & 0.62 & 1.00 \\ 
		$\sigma_{\epsilon^2_{14}}$ & 0.23 & 0.01 & 0.22 & 0.25 & 1.00 \\ 
		$\sigma_{\epsilon^2_{15}}$ & 0.21 & 0.01 & 0.20 & 0.23 & 1.00 \\ 
		$\sigma_{\epsilon^2_{16}}$ & 0.16 & 0.01 & 0.14 & 0.18 & 1.02 \\ 
		$\sigma_{\epsilon^2_{17}}$ & 0.61 & 0.02 & 0.57 & 0.66 & 1.01 \\ 
		$\sigma_{\epsilon^2_{18}}$ & 0.27 & 0.01 & 0.25 & 0.29 & 1.00 \\ 
		$\sigma_{\epsilon^2_{19}}$ & 0.35 & 0.01 & 0.32 & 0.38 & 1.00 \\ 
		$\sigma_{\epsilon^2_{110}}$ & 0.27 & 0.01 & 0.25 & 0.30 & 1.00 \\ 
		$\sigma_{\epsilon^2_{111}}$ & 0.22 & 0.01 & 0.20 & 0.24 & 1.00 \\ 
		$\sigma_{\epsilon^2_{112}}$ & 0.35 & 0.02 & 0.32 & 0.39 & 1.01 \\ 
		$\sigma_{\epsilon^2_{113}}$ & 0.53 & 0.02 & 0.49 & 0.57 & 1.00 \\ 
		$\sigma_{\epsilon^2_{114}}$ & 0.72 & 0.03 & 0.67 & 0.77 & 1.00 \\ 
		$\sigma_{\epsilon^2_{115}}$ & 0.52 & 0.02 & 0.49 & 0.56 & 1.00 \\ 
		$\sigma_{\epsilon^2_{116}}$ & 0.26 & 0.01 & 0.23 & 0.28 & 1.00 \\ 
		$\sigma_{\epsilon^2_{117}}$ & 0.53 & 0.02 & 0.50 & 0.57 & 1.00 \\ 
		$\sigma_{\epsilon^2_{21}}$ & 0.82 & 0.10 & 0.64 & 1.04 & 1.00 \\ 
		$\sigma_{\epsilon^2_{22}}$ & 0.73 & 0.09 & 0.56 & 0.93 & 1.00 \\ 
		$\sigma_{\epsilon^2_{23}}$ & 0.92 & 0.11 & 0.72 & 1.16 & 1.00 \\ 
		\hline
	\end{tabular}
\end{table}

\newpage
\section{Population definition for the simulation study}

\begin{figure}
	\centering
	\includegraphics[angle=270,width=.8\textwidth]{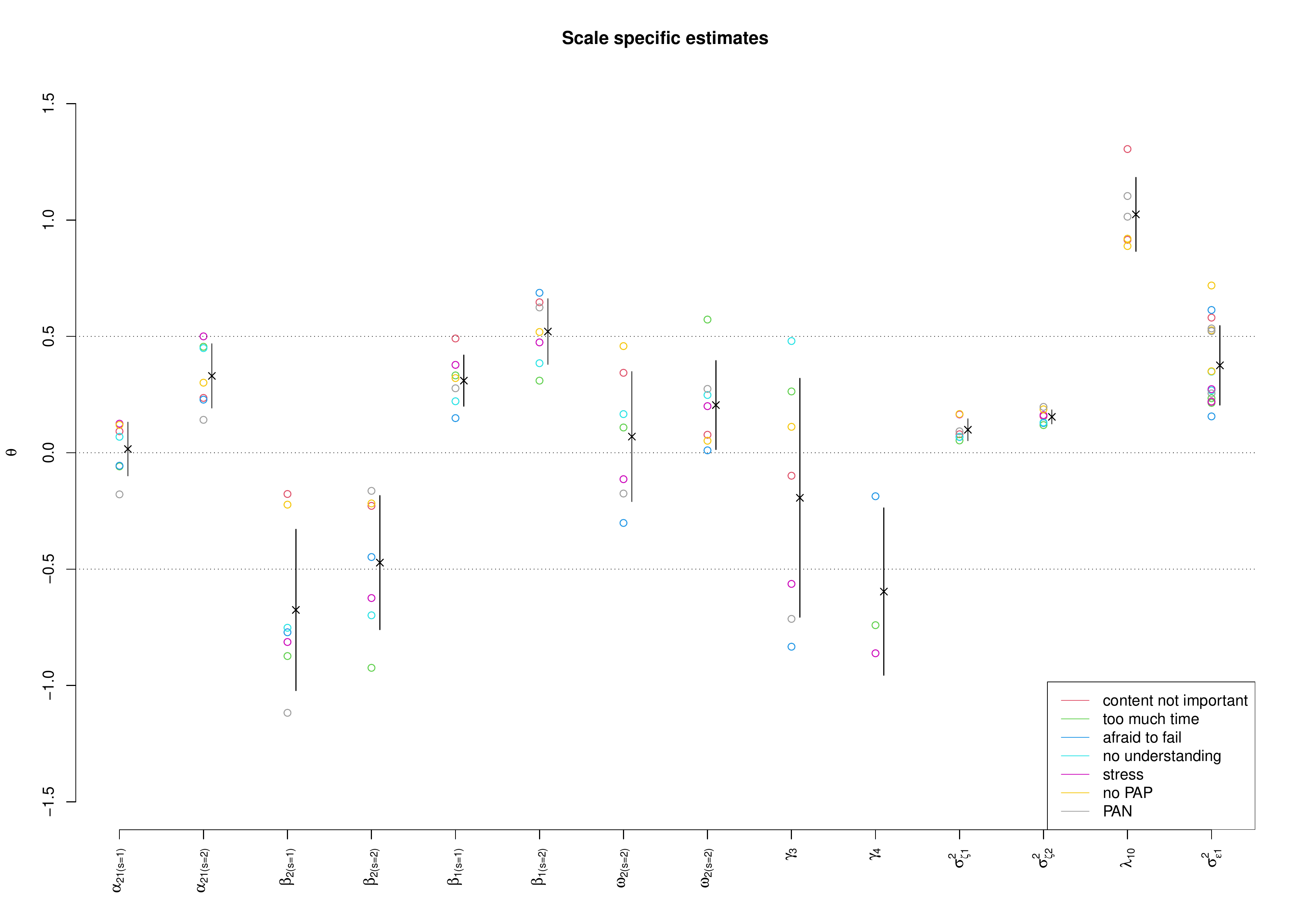}
	\caption{Scale specific parameter estimates from the empirical example. Means and standard deviations for each parameter across scales are indicated with crosses and lines.}
	\label{fig:simdata}
\end{figure}

\begin{table}[ht]
	\centering
	\caption{Population level parameters: Means and standard deviations (SD) from the empirical example.}
	\begin{tabular}{lcc}
		& & \\
		\hline
		Parameter & Mean & SD \\
		\hline
		$\lambda_{10}$ & 1.02 & 0.16 \\ 
		$\lambda_{2}$ & 0.80 & 0.56 \\ 
		$\beta_{1(s=1)}$ & 0.31 & 0.11 \\ 
		$\beta_{1(s=2)}$ & 0.52 & 0.14 \\ 
		$\beta_{2(s=1)}$ & -0.68 & 0.35 \\ 
		$\beta_{2(s=2)}$ & -0.47 & 0.29 \\ 
		$\alpha_{21(s=1)}$ & 0.02 & 0.12 \\ 
		$\alpha_{21(s=2)}$ & 0.33 & 0.14 \\ 
		$\omega_{2(s=1)}$ & 0.07 & 0.28 \\ 
		$\omega_{2(s=2)}$ & 0.20 & 0.19 \\ 
		$\gamma_1$ & 1.48 & 0.05 \\ 
		$\gamma_2$ & -0.19 & 0.51 \\ 
		$\gamma_3$ & -0.60 & 0.36 \\ 
		$\gamma_4$ & -0.71 & 0.05 \\ 
		$\sigma_{\zeta_{1j}}^2$ & 0.09 & 0.05 \\ 
		$\sigma_{\zeta_{2j}}^2$ & 0.14 & 0.03 \\ 
		$\sigma_{\zeta_{3}}^2$ & 0.20 & 0.05 \\ 
		$\sigma_{\epsilon_1}^2$ & 0.38 & 0.17 \\ 
		$\sigma_{\epsilon_2}^2$ & 0.82 & 0.10 \\ 
		\hline
		\hline
	\end{tabular}
	\label{tab:simdata}
\end{table}

\end{document}